\documentclass[fleqn,usenatbib,useAMS]{mnras}

\usepackage{newtxtext,newtxmath}

\usepackage[T1]{fontenc}
\usepackage{ae,aecompl}

\usepackage{natbib}

\usepackage{graphicx}	
\graphicspath{{figures/}}
\usepackage{amsmath}	
\usepackage{amssymb}	

\def\reff@jnl#1{{\rm#1\/}}

\def\aj{\reff@jnl{AJ}}                  
\def\araa{\reff@jnl{ARA\&A}}            
\def\apj{\reff@jnl{ApJ}}                
\def\apjl{\reff@jnl{ApJ}}               
\def\apjs{\reff@jnl{ApJS}}              
\def\apss{\reff@jnl{Ap\&SS}}            
\def\aap{\reff@jnl{A\&A}}               
\def\aapr{\reff@jnl{A\&A~Rev.}}         
\def\aaps{\reff@jnl{A\&AS}}             
\def\baas{\reff@jnl{BAAS}}              
\def\jrasc{\reff@jnl{JRASC}}            
\def\memras{\reff@jnl{MmRAS}}           
\def\mnras{\reff@jnl{MNRAS}}            
\def\physrep{\reff@jnl{Phys.Rep.}}
\def\pra{\reff@jnl{Phys.Rev.A}}         
\def\prb{\reff@jnl{Phys.Rev.B}}         
\def\prc{\reff@jnl{Phys.Rev.C}}         
\def\prd{\reff@jnl{Phys.Rev.D}}         
\def\prl{\reff@jnl{Phys.Rev.Lett}}      
\def\pasp{\reff@jnl{PASP}}              
\def\pasj{\reff@jnl{PASJ}}              
\def\skytel{\reff@jnl{S\&T}}            
\def\solphys{\reff@jnl{Solar~Phys.}}    
\def\sovast{\reff@jnl{Soviet~Ast.}}     
\def\ssr{\reff@jnl{Space~Sci.Rev.}}     
\def\nat{\reff@jnl{Nature}}             

\def\farcs{\hbox{$.\!\!^{\prime\prime}$}}

\newcommand{\beq}{\begin{equation}}
\newcommand{\eeq}{\end{equation}}
\newcommand{\beqa}{\begin{eqnarray}}
\newcommand{\eeqa}{\end{eqnarray}}

\newcommand{\rmd}{\mathrm{d}}

\newcommand{\sersic}{S\'{e}rsic}
\newcommand{\avg}[1]{\ensuremath{\left\langle #1 \right\rangle}}

\newcommand{\complexi}{\text{i}}

\newcommand{\euclid}{\emph{Euclid}}
\newcommand{\wfirst}{{\it Roman} Space Telescope}

\newcommand{\ngmix}{\textsc{ngmix}}

\newcommand{\metacal}{{\sc metacalibration}}
\newcommand{\Metacal}{{\sc metacalibration}}
\newcommand{\metadet}{{\sc metadetection}}
\newcommand{\galsim}{{\sc GalSim}}
\newcommand{\ksb}{{\sc KSB}}
\newcommand{\regauss}{{\sc Regaussianisation}}

\title[Metacalibration with undersampling]{Mitigating the effects of undersampling in weak lensing shear estimation with metacalibration}

\author[A. Kannawadi, E. Rosenberg \& H. Hoekstra]{
Arun Kannawadi$^{1,2}$\thanks{E-mail: arunkannawadi@astro.princeton.edu},
Erik Rosenberg$^{3}$,
Henk Hoekstra$^{2}$
\\
$^{1}$Department of Astrophysical Sciences, Princeton University, 4 Ivy Lane, Princeton, NJ 08544, USA\\
$^{2}$Leiden Observatory, Leiden University, PO Box 9513, 2300 RA, Leiden, the Netherlands\\
$^{3}$Institute of Astronomy, University of Cambridge, Madingley Road, Cambridge CB3 0HA, UK\\
}

\date{Accepted 2021 January 21. Received 2021 January 6; in original form 2020 October 8}

\pubyear{2021}
\begin{document}
\bibliographystyle{mnras}
\label{firstpage}
\pagerange{\pageref{firstpage}--\pageref{lastpage}}
\maketitle

\begin{abstract}
\Metacal\ is a state-of-the-art technique for measuring weak gravitational lensing shear from well-sampled galaxy images. We investigate the accuracy of shear measured with \metacal\ from fitting elliptical Gaussians to undersampled galaxy images. In this case, \metacal\ introduces aliasing effects leading to an ensemble multiplicative shear bias about $0.01$ for \euclid\, and even larger for the \wfirst, well exceeding the missions' requirements. We find that this aliasing bias can be mitigated by computing shapes from weighted moments with wider Gaussians as weight functions, thereby trading bias for a slight increase in variance of the measurements. We show that this approach is robust to the point-spread function in consideration and meets the stringent requirements of \euclid\ for galaxies with moderate to high signal-to-noise ratios. We therefore advocate \metacal\ as a viable shear measurement option for weak lensing from upcoming space missions.
\end{abstract}




\section{Introduction}
\label{sec:introduction}
Weak gravitational lensing is a powerful tool to study the growth of large-scale structure in the Universe~\citep[see][for reviews]{HoekstraJain2008, Kilbinger15,Mandelbaum2018a}. The shearing of the images of distant galaxies by the gravitational tidal field of intervening structures is now routinely measured in ever larger surveys \citep[e.g.][]{DES,deJong2013,Aihara2018}. The statistics of the resulting correlations in galaxy shapes can be directly compared to models of structure formation, thus constraining cosmological parameters \citep[e.g.][]{Troxel2018,Hikage2019,Asgari2021}. The precision will improve dramatically in the next decade with the commencement of a number of large surveys, referred to as Stage IV surveys by the Dark Energy Task Force~\citep{Albrecht2006}. Of these, \euclid\footnote{\url{https://www.euclid-ec.org}}~\citep{Laureijs2011} and the {\it Nancy Grace Roman} Space Telescope\footnote{\url{https://roman.gsfc.nasa.gov/}}~\cite[formerly known as WFIRST; ][]{spergel2015wide,Akeson2019} are space-based missions, whereas the Rubin Observatory Legacy Survey of Space and Time\footnote{\url{https://www.lsst.org/}}~\citep[LSST;][]{LSST_Book,Ivezic2019} will carry out observations using a new 8m Simonyi Survey Telescope on the ground. Importantly, these experiments are designed with gravitational lensing as a primary objective, so that various systematic errors could be addressed at the early stages through optimal survey design and observational strategies.

Several factors complicate the extraction of the cosmological lensing signal from observed galaxy images, but the most dominant one is the smearing of the images by the point-spread function (PSF), which distorts the shape of the galaxies and blends galaxies along similar lines of sight that may carry different shear signals. 
Correcting for the effects of the PSF, even for isolated galaxies, in the presence of pixel noise leads to biased estimates of galaxy shapes, and hence to biased estimates of shear \citep[e.g.][]{HS03}. Quantifying and correcting for these biases accurately is still an area of active research. Numerous algorithms, of several flavours, have been developed to estimate shear with minimal biases. Broadly speaking, there are two classes of shape measurement methods \citep[but see][which unifies them in the absence of blending]{Simon2017}. 
\begin{enumerate}
    \item Model-fitting methods assume a parametric form for the surface brightness profile. The model is convolved with the PSF and the best fit to the data is determined. Various implementations have been developed, most notably {\sc lensfit}~\citep{Miller2007,Miller2013}, {\sc im3shape}~\citep{Zuntz13}, and \ngmix\footnote{\url{https://github.com/esheldon/ngmix}}~\citep{Sheldon2015}.
    \item Moment-based methods measure a few low-order moments of the images, with a compact weight function centred at each source (see equation~\ref{eq:quad_moments}). The observed quadrupole (second-order) moments are corrected for bias introduced by the PSF and the weight function using higher order moments. Examples of this approach are \ksb~\citep{KSB95,Luppino1997} and \regauss~\citep{HS03}.
\end{enumerate} A comprehensive list of the methods along with their baseline performances can be found in~\cite{GREAT3_results1}. The performance of a shape measurement algorithm is typically assessed using simulated images of galaxies~\citep[see][for a recent review]{Plazas2020}. Through such simulations, a series of community-wide blind challenges have provided important insights into the factors that affect the performance of these algorithms~\citep{GREAT08,Kitching2011,GREAT3}. As shown by \cite{Hoekstra2015} and explored further in \cite{Hoekstra2017}, the inferred bias depends on the input parameters of the image simulations, such as the number density of galaxies and stars beyond the detection limit, but also the morphological parameters of the bright input galaxies. The presence of undetected faint sources also changes the noise properties, thereby affecting the galaxy shapes~\citep{Gurvich2016, Eckert2020}. Thus, inferring the actual shear bias in any observed data set requires dedicated image simulations that match the data as closely as possible, both in terms of galaxy properties and observational conditions~\citep[][]{FenechConti2017,Mandelbaum2018,Zuntz2018,Kannawadi2019}. These image simulations currently rely on data from the {\it Hubble} Space Telescope for obtaining the parent galaxy catalogue, which may become too restricted in terms of depth and cosmic variance~\citep{Kannawadi2015} to use to calibrate shear for Stage IV surveys. Furthermore, these dedicated simulations need to capture the clustering of galaxies at the required depth to evaluate the bias from blending~\citep{Samuroff2018, Martinet2019, Kannawadi2019}. Since most shear measurement methods exhibit biases that are a function of galaxy properties, any uncertainty in the true galaxy population, particularly at the faint end, translates to an uncertainty in shear calibration, which may be the dominant contribution to the overall error budget~\citep[e.g.][]{Hildebrandt2020}.

Recently,~\cite{Huff2017} and~\cite{Sheldon2017} proposed a method known as \Metacal\ (cf. Section~\ref{sec:metacal}), that is sufficiently unbiased for any (shear independent) subset of galaxies. Following a similar proposal earlier by~\cite{Kaiser2000}, \metacal\ estimates the bias in the shear measurement (using any shear measurement algorithm) from the data themselves.  This method has been validated on image simulations mimicking ground-based seeing conditions to subper cent accuracy in shear bias~\citep{Huff2017, Sheldon2017} and has been applied to Dark Energy Survey - Year 1 data~\citep{Zuntz2018}. The cosmological constraints obtained in~\cite{Troxel2018} from a galaxy shape catalogue measured with \textsc{im3shape} and calibrated with image simulations are consistent with those obtained from a shape catalogue measured with \ngmix\, calibrated on the data with \metacal. This demonstrates the ability of the \metacal\ algorithm to calibrate shear to an accuracy to better than 1 per cent without the need for ultra-realistic galaxy populations in the simulations. However, image simulations are nevertheless required to validate that shear measurement satisfies the stringent requirements of the Stage IV surveys~\cite[e.g.][]{Sanchez2020}. For instance, the \euclid\ mission requirements document~\citep{Laureijs2011} mandates that the shear be recovered better than $0.2$ per cent to realise the promised figure of merit. At such increased levels of accuracy, one has to worry about systematic effects that were largely ignorable in the past. These include, but are not limited to, instrumental effects such as the brighter-fatter effect~\citep{Antilogus2014, Coulton2018}, charge transfer inefficiency~\citep[][]{Rhodes2010,Israel2015} and astrophysical effects such as the presence of colour gradients in galaxy profiles due to wavelength dependence of a diffraction limited PSF~\citep{Voigt2012,Semboloni2013,Carlsten2018,Er2018} and blending of galaxies~\citep{Dawson2016, Hoekstra2017, Martinet2019}.
An updated version of \metacal\ with object detection included~\citep[\metadet;][]{Sheldon2020} provides a way forward to account for some of the blending related biases.

A potential limitation for \metacal\ in the case of space-based missions is pixelization. While the space-based surveys enjoy a better and more temporally stable PSF than their ground-based counterparts\footnote{Even ground-based PSFs may be undersampled under extremely good seeing conditions, which ironically then have to be rejected~\cite[e.g.][]{Bosch2018}}, their PSFs are typically not Nyquist-sampled in any given exposure, despite the detectors having physically smaller pixels. This can lead to small galaxies being poorly sampled as well. As \metacal\ involves reconstructing the galaxy profile from a discretized image (cf. Section~\ref{sec:metacal}), when the galaxy image is undersampled, the interpolation step will be unable to recover the galaxy profile faithfully. In this paper, using image simulations, we investigate the level of residual biases as a result of having an undersampled PSF. In our companion paper by~\cite{Hoekstra2021}, we further validate our conclusions with more realistic simulations. While both of these papers adopt \euclid\ as a reference, our results are fairly generic and are also applicable to the High-Latitude Imaging Survey\footnote{\url{https://www.roman-hls-cosmology.space/}} (HLS), the weak lensing programme of the \wfirst\ with the aim of studying dark energy parameters.

This paper is organized as follows. In Section ~\ref{sec:overview}, we provide an overview of the \metacal\ procedure and how pixelization can degrade the performance of the procedure. We explain the computational methods used in this study in Section~\ref{sec:method} and show our main results on shear bias in Section~\ref{sec:results}. Finally, strategies for mitigating this bias are explored in Section~\ref{sec:mitigation} and a general discussion of our results follows in Section~\ref{sec:discussion}.

\section{Overview of Shape measurements and Metacalibration}
\label{sec:overview}
We describe the effects of pixelization in the context of quadrupole moments, since they are more amenable to analytical calculations than model parameters, although we believe both classes of methods should be affected in a similar manner for unblended sources~\citep[see][which unifies both frameworks]{Simon2017}. We will defer the explicit treatment of model-fitting methods on undersampled images to a later work.

\subsection{Shape measurements}
\label{sec:shape_measurements}
Mathematically, the quadrupole moment is a symmetric rank-2 tensor that characterizes the distribution of an object around a chosen point, usually the centroid. Despite allowing for an unbiased estimate of the shear, one cannot use unweighted moments in the presence of noise. One must use weighted moments instead, which inevitably leads to biases that have to be calibrated out. The components of the quadrupole moments\footnote{We have chosen to ignore the normalization to keep the definition of moments linear in $I(\mathbfit{x})$, since we are not interested in measurements of flux but of shapes, which are given by a ratio of moments (or their linear combinations)} are given by
\begin{equation}
    Q_{ij} = \int\rmd^2 \mathbfit{x}\,x_i x_j W(\mathbfit{x}) I(\mathbfit{x}),
    \label{eq:quad_moments}
\end{equation}
where $x_i$ and $x_j$ are the components of $\mathbfit{x}$ for $i,j \in \lbrace 1,2 \rbrace$ with its centroid at the origin and $W(\mathbfit{x})$ could in principle be an arbitrary non-negative function that is sufficiently localized around $\mathbfit{x}=0$ to suppress the noise at large distances.~\cite{BJ02} suggest for maximum signal-to-noise (SNR) the use of adaptive moments, where the weight function is the best-fitting elliptical Gaussian function of the galaxy itself.

The complex ellipticity in terms of unweighted quadrupole moments is then given by the combination $Q_{11}-Q_{22} + 2{\rm i} Q_{12}$, with a normalizing factor of overall size, given by either $Q_{11}+Q_{22}$ or $Q_{11}+Q_{22}+2\sqrt{Q_{11}Q_{22}-Q_{12}^2}$ so that the ellipticity is invariant under flux scaling and length scaling. Following the notation of~\citet{Bartelmann2001}, we refer to the former as $\chi$-type ellipticity (also known as \emph{distortion}) and the latter as $\epsilon$-type ellipticity, and they transform\footnote{These transformations can be obtained from the transformation properties of the unweighted quadrupole moments.} differently under a lensing shear as follows:
\begin{equation}
    \epsilon = \frac{{\epsilon}^\text{(int)} + {g}}{1 + {g}^* { \epsilon}^\text{(int)}} \approx {\epsilon}^\text{(int)} + {g} - { g}^*\left( {\epsilon}^\text{(int)} \right)^2 + \mathcal{O}(|g|^2),
\end{equation}
and
\begin{equation}
    {\chi} = \frac{{\chi}^\text{(int)} + 2{g} + {g}^2{{ \chi}^\text{(int)}}^*}
    {1+{g}^*{g} + ({g}^*{\chi}^\text{(int)}+{g}{{\chi}^\text{(int)}}^*)},
\end{equation}
where $\epsilon^{\text{(int)}}$ and $\chi^{\text{(int)}}$ denote the intrinsic ellipticity of a galaxy and $g=g_1+{\rm i}g_2$ is another complex quantity referred to as the (reduced) lensing shear~\citep{Schneider1995,Seitz1997}. Thus, the ellipticity of a galaxy is a noisy estimate of the shear, with the intrinsic shape being the noise here. Invoking the isotropy assumption, $\avg{\epsilon^\text{(int)}} = \avg{ \chi^\text{(int)}}  = 0$, we can see that the ellipticity can be considered as a one-point estimator of shear since
\begin{equation}
    \avg{{\epsilon}} = g
\end{equation}
and
\begin{align}
    \avg{{\chi}} &= {\left(2-\avg{\chi^\text{(int)}{\chi^\text{(int)}}^*} \right)}{g} + \mathcal{O}({g}|{g}|^2).
    \label{eq:distortion}
\end{align}
Thus, at least to the lowest order, the ellipticity of a galaxy is a noisy estimate of the shear, up to a multiplicative factor. In practice, the dispersion in the intrinsic ellipticity is replaced by that of the measured ellipticity and the equation is still accurate to the lowest order.

The quadrupole moments as defined in equation~(\ref{eq:quad_moments}) are themselves linear in the pixel values (this is not quite true for adaptive moments), but the ellipticity is a nonlinear function of pixel values, which can lead to biased measurements in the presence of pixel noise~\citep{Kacprzak2012, Melchior2012}. Moreover, the cosmological shear signal is captured by the ellipticity of the galaxy profile, but the observed image of the galaxy is smeared by the PSF which has to be corrected for as well. Even with a perfect knowledge of the PSF, the PSF-correction is only approximate (and hence imperfect) due to its perturbative nature, acting as yet another source of bias. 

For small values of shear, which is usually the case in weak gravitational lensing, we can capture the deviation of the estimated shear $\hat{\mathbfit{g}}$ from the true shear $\mathbfit{g}_\text{true}$ from the linear bias model~\citep{STEP1,Massey2007} with a two-component additive bias $\mathbfit{c}$ and a $2\times 2$ multiplicative bias tensor $\mathbfss{M}$ as
\begin{equation}
    \hat{\mathbfit{g}} - \mathbfit{g}_\text{true} = \mathbfss{M}\mathbfit{g}_\text{true} + \mathbfit{c},
    \label{eq:linear_bias_model}
\end{equation}
with
\begin{align}
\mathbfit{c} = (c_1,c_2)^T,
\end{align}
and
\begin{align}
\mathbfss{M} = \begin{pmatrix} m_1 & m_{12} \\ m_{21} & m_2\end{pmatrix}.
\end{align}
Typically, the off-diagonal elements $m_{12}$ and $m_{21}$ are negligible and $m_{1} \approx m_{2}$. Thus, $\mathbfss{M}$ is treated as a rotationally invariant scalar in practice. However, we will use a more generic $2 \times 2$ tensor in this study. We also denote the shear in bold font as a two-component spinor instead of a complex number for convenience.

\subsection{Metacalibration algorithm}
\label{sec:metacal}
The multiplicative bias may be seen as an average response of the shapes of a population of galaxies to a coherent shear.
The \metacal\ algorithm attempts to estimate empirically the responsitivity of each galaxy in the data. It involves deconvolving the observed image by the PSF $P$ to obtain the (noisy) galaxy profile, shearing the galaxy by a small amount ($|\mathbfit{g}| \sim 0.01$) and reconvolving by a slightly larger PSF given by $\Gamma(\mathbfit{x}) = P\left[(1+2|\mathbfit{g}|)^{-1}\mathbfit{x}\right]$ to obtain the galaxy image under a different shear condition. Mathematically, we can describe the newly obtained image as the inverse Fourier transform of
\begin{equation}
\tilde{I'}(\mathbfit{k} \rvert \mathbfit{g}) = \tilde{\Gamma}^*(\mathbfit{k}) \hat{s}_\mathbfit{g} \left( \frac{\tilde{I}(\mathbfit{k})}{\tilde{P}^*(\mathbfit{k})} \right),
\label{eq:metacal_shear}
\end{equation}
where $\hat{s}_\mathbfit{g}$ is the shearing operator and $\tilde{f}(\mathbfit{k})$ stands for the Fourier transform of any real-space function $f(\mathbfit{x})$.
The per-galaxy responsitivity to shear, a $2 \times 2$ tensor $\mathbfss{R}$ defined as
\begin{equation}
    \mathbfss{R} := \frac{\partial \mathbfit{e}}{\partial \mathbfit{g}}\biggr\rvert_{\mathbfit{g}=0} \approx \frac{\mathbfit{e}^+ - \mathbfit{e}^-}{ \mathbfit{g}^+ - \mathbfit{g}^-},
\end{equation}
where $\mathbfit{g}^+$ and $\mathbfit{g}^-$ are small artificial (stimuli) shears applied to the deconvolved image through the shearing operator $\hat{s}_\mathbfit{g}$, and $\mathbfit{e}^+$ and $\mathbfit{e}^-$ are the corresponding measured ellipticities. Here, we simply denote the ellipticity in spinor form by $\mathbfit{e}$ and do not make a distinction between the $\epsilon$- and $\chi$-type ellipticities as the multiplicative factor is naturally captured by the shear response. The shear responsitivity includes much of the raw multiplicative bias in the estimator 
and the unbiased shear estimate in the absence of PSF anisotropy is
\begin{equation}
    \avg{\mathbfit{g}} \approx \avg{\mathbfss{R}}^{-1}\avg{\mathbfit{e}}.
    \label{eq:shear_estimator}
\end{equation}

Note that in this estimator, all the galaxies carry equal weight and it ignores the uncertainty in shape estimates. Depending on the level of noise, we might want to downweight some galaxies in comparison to others but in a manner that does not depend on the ellipticity itself. The generalization of the shear estimator given in equation~(\ref{eq:shear_estimator}) in the presence of galaxy weights $w$ is
\begin{equation}
    \hat{\mathbfit{g}} = \avg{w\mathbfss{R}}^{-1}\avg{w \mathbfit{e}}.
    \label{eq:generalized_shear_estimator}
\end{equation}

\subsection{Effects of pixelization}
\label{sec:pixelization}
The effect of sampling and aliasing in the images can be best understood in the Fourier domain. In the absence of noise and sky background, the observed image is the inverse Fourier transform of $\tilde{I}(\mathbfit{k}) = \tilde{G}(\mathbfit{k})\tilde{P}(\mathbfit{k})$, where $\tilde{G}(\mathbfit{k})$ and $\tilde{P}(\mathbfit{k})$ are the Fourier transforms of the galaxy and the PSF respectively. In practice, $\tilde{I}(\mathbfit{k})$ has to be computed at discrete values of $\mathbfit{k}$ from a discrete (pixellated) image of finite extent. The \galsim\ implementation of the \metacal\ algorithm involves computing a discrete approximation to the Fourier transforms by first constructing a continuous function by interpolating the pixel values, and taking the Fourier transform assuming a periodic boundary condition, which is arguably better than using discrete Fourier transforms.
 
Note that the quantities above, especially $\tilde{I}(\mathbfit{k})$, are not discrete Fourier transforms of the pixelized image but a discrete approximation to the Fourier transform of $I^\text{interp}(\mathbfit{x})$ obtained by interpolating from the discrete image. Due to the diffraction limit, there exists a $k_{\max}$ such that $\tilde{P}(\mathbfit{k}) \equiv 0$ for all $\mathbfit{k}$ with magnitude $|\mathbfit{k}|>k_{\max}$. This holds true irrespective of the central obscuration in the telescope aperture. Thus, ignoring the read noise, the observed image is band-limited. For interpolation using the sinc function, which is perfect for band-limited signals, the interpolated image is formally the inverse Fourier transform of
\begin{equation}
    \tilde{I}^\text{interp}(\mathbfit{k}) = \left[ \sum_{n_1=-\infty}^{\infty}\sum_{n_2=-\infty}^{\infty}\tilde{I}(\mathbfit{k}+\mathbfit{n}\Delta_{\mathbfit{k}}) \right] \Theta(\Delta_{\mathbfit{k}}-|k_1|)\Theta(\Delta_{\mathbfit{k}}-|k_2|),
    \label{eq:I_interp}
\end{equation}
where $\mathbfit{n}=(n_1,n_2)$, $\Theta(\cdot)$ is the Heaviside step function, and $\Delta_{\mathbfit{k}}$ is the periodicity in Fourier space due to starting from discrete pixel values. For square pixels of physical size $a$, $\Delta_{\mathbfit{k}} = 2\upi/a$. 

According to the Nyquist-Shannon sampling theorem, in order to reconstruct a continuous function from discrete samples by interpolation, the (spatial) sampling frequency has to be more than twice the maximum frequency present in the original signal, also referred to as the Nyquist frequency. This follows from equation~(\ref{eq:I_interp}): if $\tilde{I}(\mathbfit{k}) = 0$ for $|\mathbfit{k}|>\frac{1}{2}\Delta_{\mathbfit{k}}$, then the only contributing term is that of $\mathbfit{n}=0$ and $\tilde{I}^\text{interp}(\mathbfit{k}) \equiv \tilde{I}(\mathbfit{k})$. If the above condition is not satisfied, then frequencies higher than the Nyquist frequencies are misinterpreted as lower frequencies, a phenomenon well-known as \emph{aliasing}. If we perform operations such as applying a lensing shear, the images obtained by shearing $I^\text{interp}$ would be systematically different from that obtained from $\tilde{I}(\mathbfit{k})$, were it to be accessible.

We define\footnote{Our definition differs from that of~\cite{IMCOM_WLsystematics, Plazas2016} by a factor of two.} the sampling factor $Q$ as the ratio of the Nyquist scale (reciprocal of Nyquist frequency) to the pixel scale. An image is said to be Nyquist-sampled if $Q > 1$, critically sampled if $Q = 1$ and undersampled if $Q<1$. An undersampled image is said to be strongly undersampled if $Q<0.5$, where all frequency modes are aliased. Incidentally, if $Q<0.5$, the true profile of the object cannot be recovered from any set of four dithered exposures~\citep[see][for example]{Lauer99}.

If the PSF is Nyquist-sampled then by definition, $\Delta_{\mathbfit{k}} > 2k_{\max}$ for all galaxies and there is no signal present if $\mathbfit{n} \ne 0$. $I^\text{interp}(\mathbfit{x})$ has only a small contribution only from the (subdominant) detector noise, whose effect is to change the inferred property of the noise characteristics. For an undersampled PSF, in addition to the detector noise, there is a finite contribution from the signal as well. This results in $\tilde{I}'(\mathbfit{k}|\mathbfit{g})$ in equation~(\ref{eq:metacal_shear})  being different from $\tilde{I}(\mathbfit{k})$ even when $\mathbfit{g}=0$. While the PSF is not Nyquist-sampled, it can nevertheless be modelled very well from multiple star images, so we will assume that an oversampled image of the PSF is available which does not suffer from aliasing when interpolated between pixels. In this paper, we refer to the bias in shear estimate arising due to aliasing as `aliasing bias'. We refrain from referring to it as `pixelization bias', in order to avoid confusing this with other sources of bias such as discretisation of a continuous image.

A distinctive signature of aliasing is that it affects $g_1$ preferentially more than $g_2$. We motivate this generic result using symmetry. Aliasing introduces an oscillatory pattern along the axes, modulated by the image $I(\mathbfit{x})$. If, on average, the image is isotropic but for a small $g_2$, the contributions from the $\mathbfit{n}\ne 0$ term in the integral for $Q_{12}$ are zero. However, if there is a net $g_1$, then the horizontal and vertical axes are not treated on the same footing. The oscillations do not cancel each other, contributing to $Q_{11}-Q_{22}$. This leads to a non-zero $m_1$, but not $m_2$.
In an image with discrete pixels, the subpixel offset breaks this symmetry, which leads to a non-zero but small $m_2$ in practice.

\section{Method}
\label{sec:method}
Our goal is to evaluate the impact that aliasing would have for measuring shear using \metacal\ from planned Stage IV surveys. Specifically, we study the performance of \metacal\ on simulated galaxy images, some of which may be poorly sampled. We use the specifications of \euclid\ in our fiducial simulation set-up, and extend conclusions to the \wfirst\ by considering corresponding sampling factors. The \euclid\ telescope has an entrance pupil that is 1.2\,m in diameter. \euclid\ will cover about $15\,000$ deg$^2$ of the sky, thereby imaging about two billion galaxies with an SNR greater than $10$. The images for shape measurements are captured using the \euclid\ visible (VIS) instrument\footnote{\url{ https://sci.esa.int/s/w7dEOxW}}, with a single broad band filter covering a wide range of wavelengths from 550\,nm to 900\,nm and at a resolution of 0\farcs{1} pixel$^{-1}$. The PSF is wavelength dependent set by the diffraction limit, and as a result, each galaxy is convolved by a different PSF determined by its spectral energy distribution (SED).

\subsection{Image simulations}
\label{sec:simulations}
\begin{figure}
    \centering
    \includegraphics[width=\columnwidth]{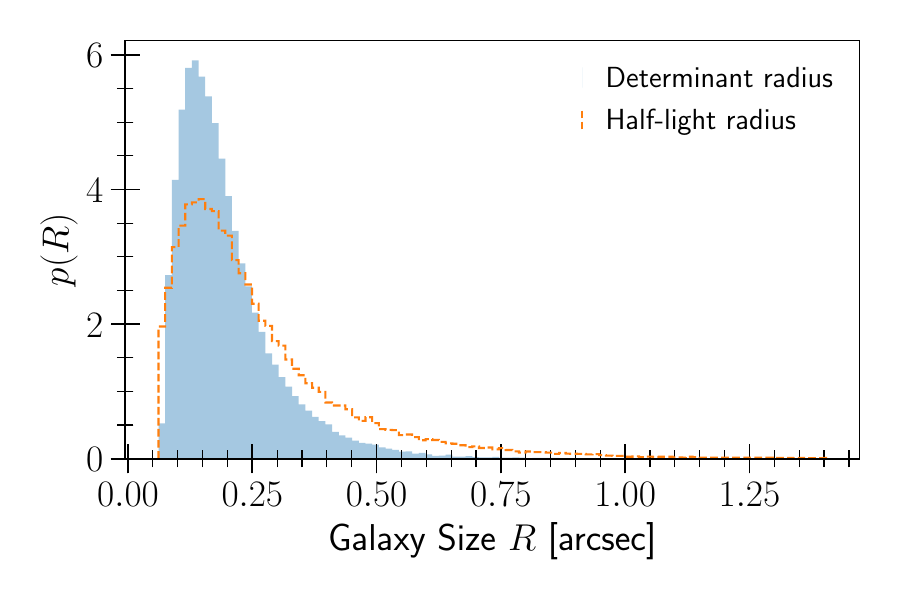}
    \caption{Histogram of two galaxy size measures: circularized half-light radius and determinant radius (fourth root of determinant of the moment matrix) of the best-fitting elliptical Gaussian measured using adaptive moments. Both these measures are azimuthally averaged.}
    \label{fig:size_hist}
\end{figure}
Postage stamp images of isolated galaxies are simulated with the publicly available \galsim\footnote{\url{https://github.com/GalSim-developers/GalSim}} package~\citep{galsim}. We model the surface brightness profiles of galaxies using \sersic\ profiles~\citep{Sersic1968}, which have well-defined ellipticities. \galsim\ comes equipped with a \texttt{COSMOSCatalog}, which is a catalogue of best-fitting \sersic\ parameters to galaxies in the COSMOS survey~\citep{COSMOS_generic,COSMOS_Alexie,COSMOS_overview}. The best-fitting parameters were obtained using the method described in~\cite{Claire_Fits}. This serves as our parent galaxy catalogue in estimating shear biases (cf. Section~\ref{sec:results}). The distribution of the sizes of the galaxies is shown in Figure~\ref{fig:size_hist}. Distributions of other galaxy properties such as apparent magnitudes, best-fitting \sersic\ parameters can be found in figs 11-13 in the appendix of~\cite{GREAT3}. Our parent galaxy catalogue is likely to be different from catalogues from \euclid\ or \wfirst\ due to different selection functions. Hence, the shear biases derived in this study are not directly applicable to those surveys but are only indicative.

Since the PSF is diffraction-limited, its size depends on the colour of the galaxy itself. For the sake of simplicity, we use a monochromatic Airy function as the PSF for all galaxies, corresponding to a wavelength of $800$\,nm (unless otherwise mentioned) and arising from a circular aperture of diameter $1.2$\,m, with a central obscuration factor of $0.3$. We exploit the perfect knowledge of this input PSF, assuming that the PSFs in the real data can be modelled to sufficient accuracy. An imperfect knowledge of the PSF will introduce bias for all shear measurement methods, and quantifying this for \metacal\ is outside the scope of this paper.

\euclid\ will image a given field using the VIS instrument with three or four dithered exposures, and the HLS with the \wfirst\ will include several dithered visits, over multiple passbands however. Therefore, we render four individual exposures for each galaxy. The exposures have the same PSF but differ by uniform subpixel offsets of half a pixel in each direction. The galaxy itself has a random subpixel shift to avoid consistently sampling at the peak of the light profile. We also include a $90^\circ$-rotated copy (about the same randomly offset centre) for each galaxy to beat down the intrinsic shape noise~\citep{Massey2007}. The mapping between sky coordinates and pixel coordinates is linear, given by a constant pixel scale of $0\farcs{1}$ pixel$^{-1}$. Since the goal of this paper is to study the bias just due to pixelization and having undersampled PSFs, we do not add any pixel noise to the images for most simulations in this paper. The rationale is that it is sufficient to show on noiseless simulations that aliasing bias is significant. Adding pixel noise complicates the study by inducing correlations in the noise and also contaminates the estimate of aliasing bias with noise bias.

Furthermore, we do not expect any significant coupling between aliasing and other detector-level systematics. The quantities $\tilde{I}^\text{interp}(\mathbfit{k})$ and $\tilde{I}(\mathbfit{k})$ (see equation~\ref{eq:I_interp}) differ systematically only for those values of $\mathbfit{k}$ for which $\Delta_{\mathbfit{k}}$ < |\mathbfit{k}| < $k_{\max}$. Instrumental effects that mix the different $k$-scales can transfer the spurious power in these high-frequency modes to lower frequencies, which, if not corrected for accurately, could bias the flux and shape measurements. However, we expect such imperfect corrections to bias the measurements regardless of whether the PSF is Nyquist-sampled or not. Hence, our simulations also assume ideal detector behaviour.

\subsection{Metacalibration procedure}
\label{sec:metacal_procedure}
Although the PSFs in space telescopes are temporally stable, the variation of the PSF over the focal plane could mean that different exposures of the same galaxy have different PSFs. For this reason, ideally, one might want to apply the \metacal\ procedure to the individual exposures with corresponding PSFs, rather than applying it to a coadded image. For convenience, we refer to the output images of metacalibration as metacalibrated exposures. We apply the metacalibration algorithm (see Section~\ref{sec:metacal}) to each exposure with five artificial shears of 0, $\pm 0.01$ and $\pm \complexi\, 0.01$, generating five metacalibrated exposures per original exposure.
 
To mimic the \metacal\ procedure that will be done on the real data, we do not use the perfectly known, smooth PSF profile, but render the PSF image by convolving the obscured-Airy profile with a top-hat pixel response function corresponding to a pixel scale $p = 0\farcs{1}$ pixel$^{-1}$ and oversampling the image by a factor of five. A smooth PSF model is then constructed from this oversampled PSF image, which is used for the deconvolution step. The PSF interpolated from a finite-sized image is not strictly band-limited. To suppress any spurious power at high spatial frequencies, we use the band-limited Airy PSF for the reconvolution step. Following the prescription in~\cite{Huff2017}, we dilate the analytical PSF prior to its convolution with the pixel response function and use an oversampled image of the same for shape measurements. The choice of re-convolving with a band-limited, preferably analytical, function should be made for the real data to suppress any artefacts arising from the deconvolution step.

We find that the \texttt{Quintic} interpolation scheme (the default option in \galsim) is far too inaccurate and causes large differences between the ellipticities prior and posterior to interpolation. Switching to a \texttt{sinc} interpolation kernel improves the accuracy, but is extremely slow to render the images after interpolation. We find \texttt{LanczosN} with $N=50$ as an optimal compromise between speed and accuracy and is our default interpolation kernel. See Appendix~\ref{app:interpolation} for more details on our choice of interpolation kernel.

In order to further ensure that our results are not affected due to moment calculations from poorly sampled PSF images or shape measurement failures in a particular subpopulation of galaxies, we employ a control branch that is similar to the idea proposed by~\cite{Pujol2019}. In our control branch, we skip the steps in metacalibration where the galaxy image is interpolated and deconvolved by the PSF. Instead, we include the artificial shear prior to PSF convolution, and then render the image with dilated PSF and measure its shear responsitivity. Therefore, by construction, we do not expect our control branch to exhibit aliasing bias. To avoid any subtle selection effects between the \metacal\ and control branches (some measurements  are failures only in the \metacal\ branch), we impose the two to have the same population of galaxies. Note that our pipeline does not include detection and deblending steps, as we simulate images of isolated galaxies. As a result of this simplification, we do not suffer from the object detection bias discussed in~\cite{Sheldon2020} and~\cite{Hoekstra2021}.

\subsection{PSF correction}
\label{sec:psfcorr}
In principle, \metacal\ can account for the PSF in the calculation of the per-object responsitivity without any explicit PSF-correction.~\cite{Hoekstra2021} show that \metacal\ works in a \euclid-like set-up (with constant PSF) without an explicit PSF-correction. However, in the case of variable PSFs it may prove beneficial to use some simple PSF-correction schemes to detrend some of the well-understood effects of PSF, so that the biases that need to be removed by \metacal\ are small to start with. While~\cite{Sheldon2020} show that smooth PSF variations, at the levels expected for the LSST, are not an issue, it remains unclear whether galaxy-by-galaxy PSF variation in \euclid\ due to their SEDs could be similarly handled. Hence, we employ a few simple PSF-correction methods and study their sensitivity to \metacal. We provide a brief overview of the methods used in this paper. 

\subsubsection{Gaussian fit}
This method approximates both the PSF and the galaxy as elliptical Gaussians. We use a simple single-Gaussian fitting method whereby we fit an elliptical Gaussian to the PSF and to the galaxy image. The difference in the covariance matrices of the two Gaussians gives the equivalent of unweighted moments for the intrinsic galaxy. Due to its simplistic assumptions, this method is expected to suffer from severe model bias~\citep{Bernstein2010, Voigt2010}. However, in combination with \metacal\, this has been shown to achieve residual biases within a per cent on real data~\citep{Sheldon2017,Zuntz2018}.

The fitting methodology employed above is different from those methods that fit a PSF-convolved galaxy model to the image and maximize the likelihood to find the best-fitting parameters. However,  in the absence of blending and missing pixels, model-fitting methods and moment-based methods are equivalent~\citep{Simon2017}. We will therefore not concern ourselves with the exact details of the model-fitting approach. 

\subsubsection{Regaussianisation}
\label{sec:regauss}
A natural extension of the idea above is captured in the \regauss\ method~\citep{HS03}. It is a perturbative method whereby a small non-Gaussianity in the PSF is accounted for and the intrinsic ellipticity of the galaxy is obtained from the second-order moments, with correction terms from fourth-order radial moments included~\citep[see][for details]{BJ02}. We used the \galsim\ implementation of the algorithm in this paper. 

\subsubsection{KSB}
\label{sec:ksb}
The \ksb\ method, named after the authors of~\cite{KSB95}, is one of the pioneering shape measurement methods completely based on image moments. We use the \galsim\ implementation of this method described in Appendix C of~\cite{HS03}. In this implementation, the ellipticity spinor is calculated from the quadrupole moments measured with a circular Gaussian weight function, whose size is typically (but not necessarily) matched to observed galaxy. The PSF moments are also calculated using the same weight function as the galaxy following~\citet{Hoekstra1998}. From the spatial derivatives of the weight function, polarizability tensors are calculated and are used to remove the effect of PSF anisotropy and obtain the shear.

It is worth noting that all of these methods are inexpensive computationally, both in terms of memory and computing cycles, in relation to \metacal\ (see Table~\ref{tab:runtime}). Also,
note that among the three methods listed above, only \ksb\ has a freedom in the choice of the weight function. We will utilise this freedom in Section~\ref{sec:mitigation}. 

The PSF-correction routines in \galsim\ require that both the galaxy and PSF moments are calculated from images with the same pixel scale. This limitation prevents us from using the perfectly known PSF model. The PSF moments (and moments of small galaxies) calculated from images at native pixel scale have sampling errors large enough to cause ensemble shear biases of the order of 0.01, but are greatly mitigated when the PSF image is oversampled by a factor of two or more~\cite[e.g.][]{Hoekstra2021}. To be consistent with these oversampled PSF images, we interleave the four metacalibrated exposures to obtain a high-resolution coadded image, with an effective pixel scale of $0\farcs{05}$ pixel$^{-1}$ using the \texttt{galsim.utilities.interleaveImage} routine~\citep{Kannawadi_IPC_PSF}. This step plays the role of a more complicated coaddition technique that may be used in practice and represents the best case scenario. A typical coadded image will not be sampled this well.  In addition to the increased resolution, the advantage of interleaving (as in any coaddition) is that it allows us to use shape information of galaxies from multiple images simultaneously, as in multi-epoch model fitting methods.

\section{Quantifying aliasing bias}
\label{sec:results}
For a (semi-)realistic population of galaxies taken from the COSMOS sample (see Section~\ref{sec:simulations}), we show how well we can recover the cosmological shear signal. As mentioned in Section~\ref{sec:shape_measurements}, we relate the recovered shear to the true shear via a generalized linear shear bias model given in equation~(\ref{eq:linear_bias_model}).

\begin{figure}
    \centering
    \includegraphics[width=\columnwidth]{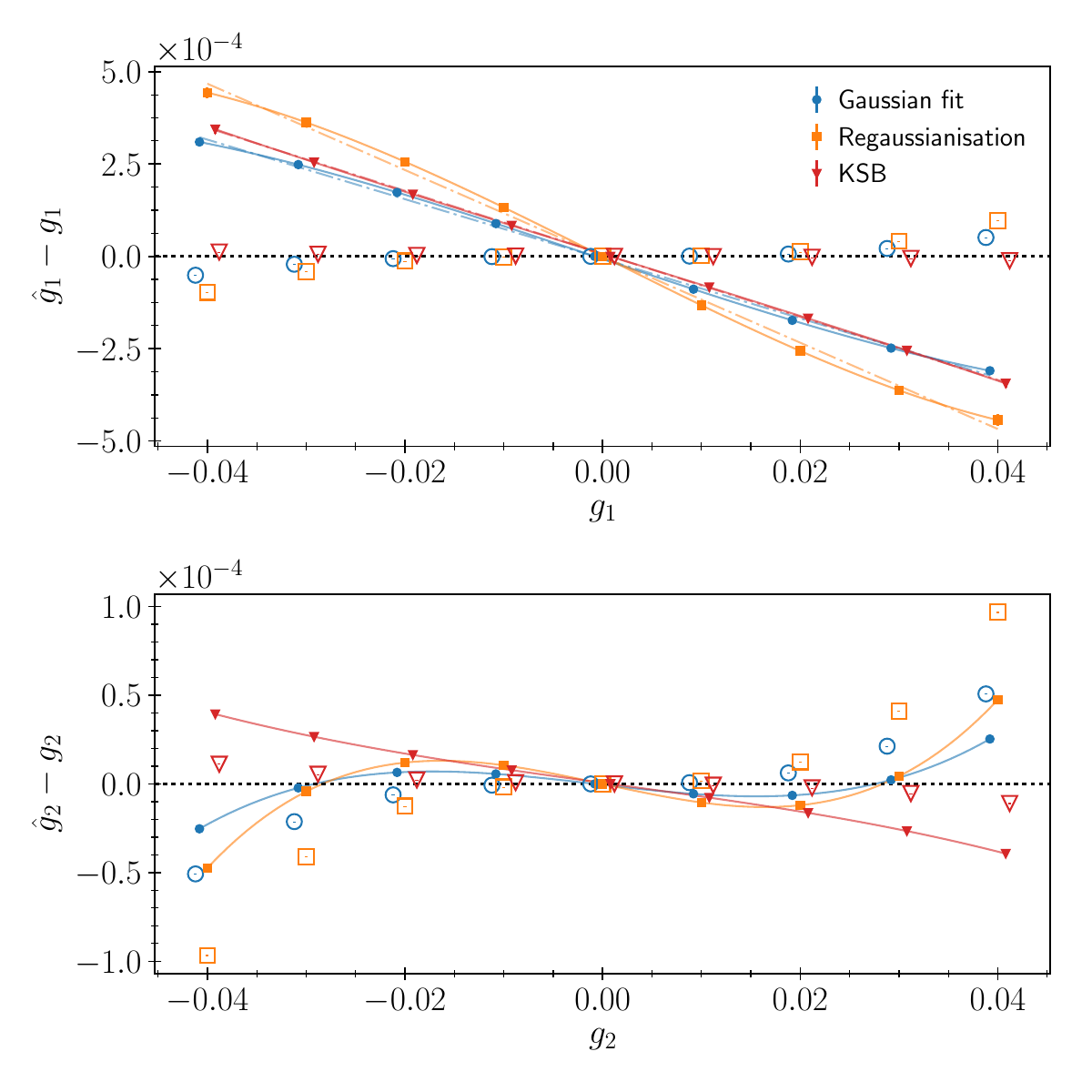}
    \caption{Relationship between input and output shears from different PSF-correction methods for a random subset of 10 000 galaxies. The true shear values range from -0.04 to 0.04 in steps of 0.01 in one component, while the other component is fixed at zero. The data points are offset slightly along the horizontal direction for clarity. The filled points are the shear recovered from the \metacal\ branch and the open points are from the control branch. The solid curves show the best-fitting cubic polynomial and the dashed-dotted lines are the best-fitting straight lines (for $g_1$ only), with the best-fitting parameters listed in Table~\ref{tab:bias_values}. }
    \label{fig:multiplicative_bias}
\end{figure}

\begin{table*}
    \caption{
    Residual multiplicative bias in shear measured with various PSF-correction methods after applying \metacal. The uncertainties capture the deviation of the mean shear from the best-fitting function, and do not include the uncertainty in the shear measurement itself. The $c$-terms are from the linear fit were very similar to cubic polynomial fit.}
    \label{tab:bias_values}
\begin{tabular}{lcrrrrr}
    \hline
    Algorithm for & Ellipticity & $m_1 [\times 10^2]$ & $m_1 [\times 10^2]$ & $m_2 [\times 10^4]$ & $c_1 [\times 10^8]$ & $c_2 [\times 10^8]$ \\
    PSF-correction & type & (linear) & (cubic) & (cubic) & (cubic) & (cubic)\\
    \hline
    Gaussian fit & $\chi$ & $-0.81 \pm 0.01$ & $-0.8968 \pm 0.0002 $ & $-6.420 \pm 0.003$ & $0.2 \pm 2.0$ & $-0.3 \pm 0.3$\\
    \regauss & $\chi$ & $-1.17 \pm 0.03$ & $-1.3380 \pm 0.0002$ & $-11.97\pm 0.01$ & $0.5 \pm 2.1$ & $-0.9 \pm 1.4$ \\
    \ksb & $\epsilon$ & $-0.853 \pm 0.003$ & $-0.8342 \pm 0.0003$ & $-7.614\pm0.002$ & $0.2 \pm 2.5$ & $0.2 \pm 0.2$\\
    \hline
\end{tabular}
\end{table*}

The primary source of the $\mathbfit{c}$ term is PSF leakage, i.e., inadequate correction for PSF ellipticity. As we have chosen a circular PSF for our study, we do not expect any additive bias terms. We checked this explicitly by seeing that we recover a null signal (about $10^{-7}$ or smaller) when the true shear is zero. Additionally, when only one of two shear components is non-zero, we recovered (almost) zero shear for the component set to zero. We also noticed a mild linear dependence on the non-zero component, corresponding to $|m_{12}| \approx |m_{21}| \lesssim 2 \times 10^{-4}$.

In practice, the PSF will be slightly anisotropic, leading to non-negligible additive bias terms due to PSF leakage, and also possibly due to pixelization effects. However, in this study, we are not interested in quantifying the contribution of pixelization to additive biases as any residual biases can be estimated and removed from the data themselves. We will therefore focus on the multiplicative bias terms, and thus the ability to recover non-zero shear. Our results already indicate that there is no discernible cross-talk between the two shear components, and therefore the multiplicative bias $\mathbfss{M}$ may be considered to be strictly diagonal.

Figure~\ref{fig:multiplicative_bias} shows the relationship between the true input shear $g_i$ and the recovered shear $\hat{g}_i$ for\,  $ i=1,2 $. The presence of a small higher order (cubic) relation is evident in the $g_2$ component, which closely follows the trend in the control branch for $g_2$ (and $g_1$). The cubic relation is a result of calculating $\chi$-type ellipticities as opposed to $\epsilon$-type ellipticities\footnote{Although the $\epsilon$-type ellipticity appears to be the preferred type as it is unaffected by higher order shear terms, the quantity under the square root is not guaranteed to be positive for PSF-corrected moments in the presence of noise, which introduces biases.} (see equation~\ref{eq:distortion}) and is inherent to the Gaussian fit and \regauss\ estimators. We notice that the systematic error in $g_1$ is dominated by a linear term, which is practically negligible in the control branch. In  Table~\ref{tab:bias_values},  we list the multiplicative and additive bias terms obtained by fitting a cubic function (without a quadratic term) to the points in Fig.~\ref{fig:multiplicative_bias}. In all cases, the value of $m_1$ is an order of magnitude larger than $m_2$ indicating that aliasing affects $m_1$ more than $m_2$, which is not surprising given the symmetry argument in Section~\ref{sec:pixelization}. We also fit a straight line to the data points in the upper panel of Fig.~\ref{fig:multiplicative_bias}, whose slopes are tabulated in Table~\ref{tab:bias_values}.  A comparison of the two $m_1$ columns for $\chi-$type measurements shows that when the nonlinear terms are dropped, the amplitude of $m_1$ is overestimated by a small amount, by approximately $m_2$. 

Since the biases from higher order shear terms, cross-terms in $\mathbfss{M}$, and $\mathbfit{c}$ are negligible, we estimate the linear multiplicative bias as
\begin{equation}
\label{eq:one_point_estimator}
    {m}_i := \hat{g}_i/g_i - 1,
\end{equation} for $i=1,2$ and for small enough $g_i$.
Note that this simple form of the estimator implies we do not have to perform a linear regression over multiple true shear values to find the multiplicative term, thereby reducing the volume of simulations required. The uncertainty in $m_i$ is then simply the uncertainty in $\hat{g}_i$ scaled by $1/|g_i|$.  In the following sections, the multiplicative bias is computed from simulations with input shear of $g_1 = g_2 = -0.02$.

\subsection{Dependence on galaxy sizes}
\begin{figure}
    \centering
    \includegraphics[width=\columnwidth]{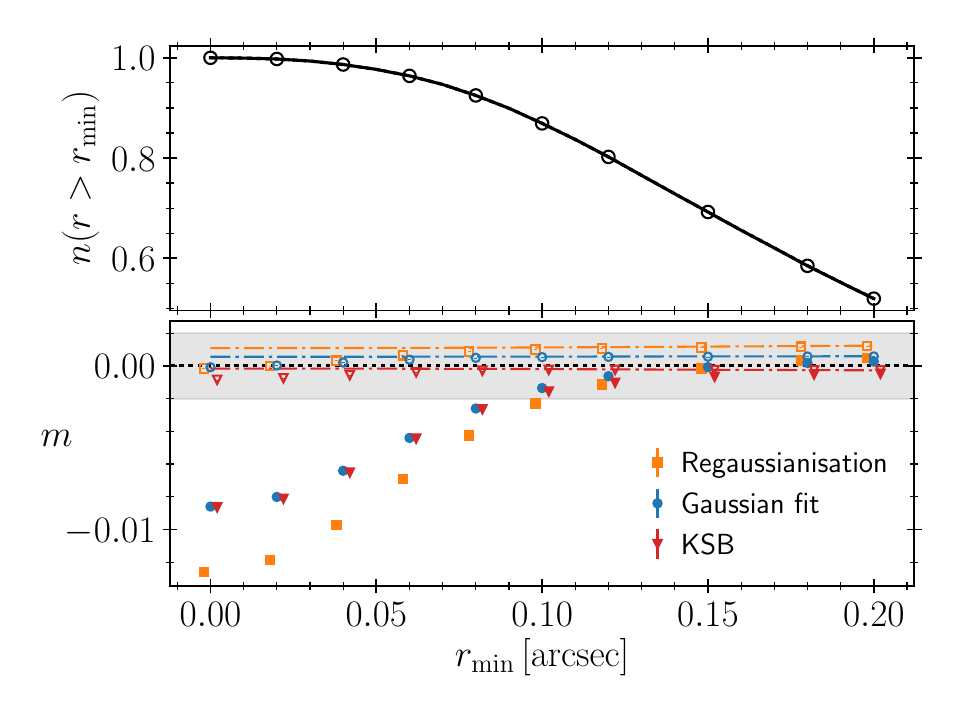}
    \caption{Lower panel: Aliasing bias as a function of minimum half-light radius imposed on the galaxy samples. The filled points show $m_1$ after \metacal\ and the open points show $m_2$. Both $m_2$ and $m_1$ for the control branch (dashed-dotted lines) show signs of nonlinear bias in $\chi$-type measurements. The shaded region represents that \euclid\ requirement on $m_1$ to be within $\pm 2\times 10^{-3}$. Upper panel: The fraction of galaxies in our sample (by number) that survive the minimum size criterion.}
    \label{fig:size_cuts}
\end{figure}

We demonstrate that the bias observed in the $g_1$ component is pre-dominantly due to a subpopulation of small galaxies, whose intrinsic sizes happen to be of the order of, or smaller than, the pixel size, and hence undersampled in a single exposure. By progressively eliminating small galaxies from our sample, we show in Fig.~\ref{fig:size_cuts} that $m_1$ converges to zero within the accepted range of accuracy. Eliminating galaxies smaller than a given threshold corresponds to a special case of equation~(\ref{eq:generalized_shear_estimator}) with a binary weight $ w(r;r_{\min}) = \Theta(r-r_{\min})$, where $\Theta(\cdot)$ is the Heaviside-step function, $r$ is the true intrinsic half-light radius of the galaxy (circularized, so that the weight is independent of ellipticity), and $r_{\min}$ is the threshold, which sets the minimum size of the galaxy in our sample. Figure~\ref{fig:size_cuts} also shows explicitly that the aliasing affects only $m_1$ and has little effect on $m_2$.

The bias for a sample of galaxies depends on its size distribution, as a result of this size dependence. The upper panel of Fig.~\ref{fig:size_cuts} shows the cumulative marginal distribution of input galaxy sizes, marginalized over other galaxy properties such as intrinsic ellipticities, \sersic\ indices etc. For the COSMOS galaxy sample we simulated, the bias is dominated by 13 per cent of the galaxies, whose intrinsic radii is smaller than $0\farcs{1}$, or 1 native pixel.

Note that we do not have a detection stage in our simplistic set-up and are also insensitive to the absolute level of surface brightness of the galaxy due to absence of a sky background and noise. While we do not apply an explicit magnitude cut in our simulations, our input catalogue is nevertheless magnitude-limited. However, those galaxies in our input catalogue with $r<0\farcs{1}$ were well resolved in the Wide Field Camera of the {\it Hubble} Space Telescope with a pixel scale of $0\farcs{05}$ pixel$^{-1}$ (and an effective pixel scale of $0\farcs{03}$ pixel$^{-1}$ after coaddition), but may escape detection or be poorly resolved in \euclid\ and \wfirst\ images which have larger pixel scales.  Thus, for future Stage IV experiments from space, an implicit selection effect due to detection may play a natural role in excluding such small galaxies and may suffer less from aliasing than shown here. On the other hand, galaxies are not smooth and have complex morphologies in reality. The size parameter $r$ merely describes how rapidly the surface brightness drops given a \sersic\ index. In practice, even large galaxies that exhibit small-scale features will contribute to aliasing (see Appendix~\ref{app:morphology} for shear measured from realistic galaxies).

A plausible explanation for the bias could be mis-centring, i.e., the offset between the centres of the galaxies and PSF models. While the PSFs are always centred at the centre of the central pixel, the galaxies have random subpixel offsets. The difference in sampling could potentially introduce biases, with smaller galaxies showing larger biases. If this was the dominant source of bias, we would then expect it to be the case even in the control branch where such mis-centring is present. Furthermore, since the PSFs are oversampled by a factor of five, it is highly unlikely that these biases could be arising due to sampling. Finally, we explicitly checked if this is the case by centering all galaxies at the pixel centre and found no significant difference. Hence, we rule out mis-centring as a potential source of the bias.

\subsection{Dependence on sampling factor}
\begin{figure}
    \centering
    \includegraphics[width=\columnwidth]{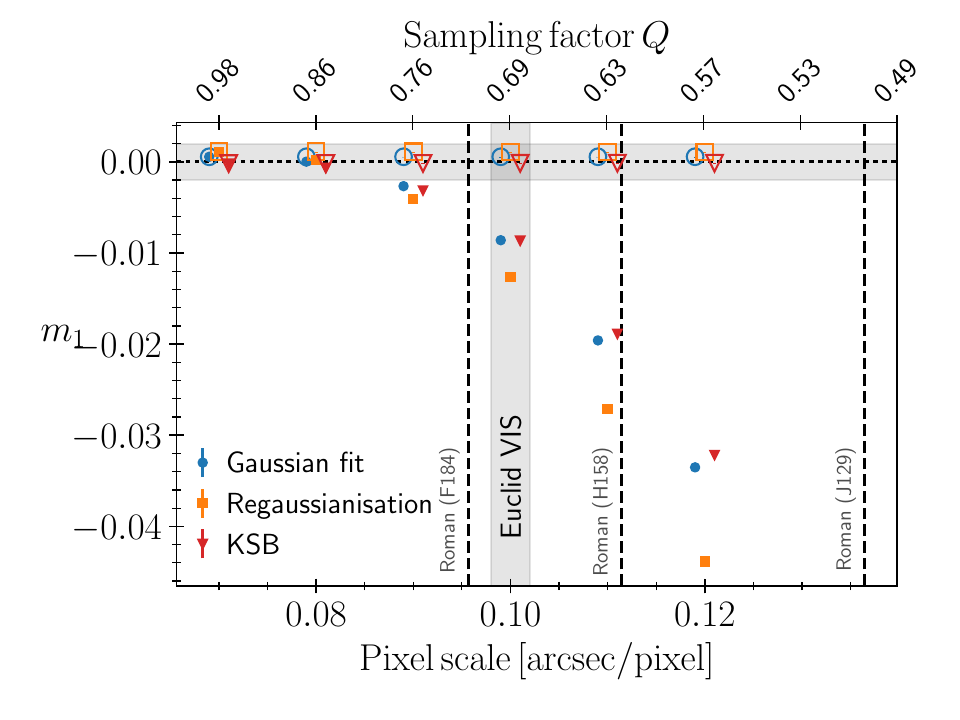}
    \caption{Plot of residual shear multiplicative bias due to pixelization for various sampling factors, achieved by varying the pixel scale. A sampling factor less than unity indicates undersampling of the PSF. The filled points show the bias after \metacal\ and the open points show the bias from the control branch. The shaded horizontal region represents the \euclid\ requirement on $m_1$ to be within $\pm 2\times 10^{-3}$ and the vertical shaded region highlights the bias incurred by various PSF-correction methods at \euclid\ pixel scale. The sampling factors for the three imaging bandpasses for the \wfirst\ are also indicated by dashed vertical lines.}
    \label{fig:money_plot}
\end{figure}

In order to establish that this bias arises from the undersampling of the images, we study how the shear is recovered as a function of sampling factor $Q$. We vary the pixel scale to vary $Q$ and show $m_1$ as a function of the pixel scale in Fig.~\ref{fig:money_plot}. The pixel scale in the \euclid\ VIS detectors is fixed at 0\farcs{1} pixel$^{-1}$ and in the \wfirst\ it is 0\farcs{11} pixel$^{-1}$. At a pixel scale of 0\farcs{1} pixel$^{-1}$ (that of \euclid\ VIS detectors), we see a clear difference in signal obtained using the \metacal\ branch and control branch. The uncertainty values are obtained from 1000 bootstrap realizations of galaxy pairs. We see that for \euclid, a residual bias in $m_1$ of approximately $0.01$ after \metacal\ is possible, while the control branch shows no biases exceeding the tolerance.

Although our set-up mimics \euclid, it is straightforward to assess the level of bias that would be incurred in a \wfirst-like setting using scaling arguments. A suite of dedicated image simulations for \wfirst, one that is more representative than this study is for \euclid, is given in~\cite{Troxel2021}, albeit without \metacal. Despite the \wfirst\ bandpasses covering the near-infrared wavelengths, the telescope aperture has twice the diameter as that of \euclid\ and the pixel scale of the H4RG detectors in \wfirst\ is only 10 per cent bigger than \euclid\ VIS detectors. As a result, the \wfirst\ PSFs are likely to be equally undersampled or worse than the \euclid\ PSFs. The sampling factor acts as a powerful summary statistic that is robust to change in the details of the PSF, pixel scale, aberrations etc.

The shape measurements for the HLS are intended to be done in three bandpass filters: $J$, $H$, $F$, whose effective wavelengths are $1290$, $1580$ and $1840$\,nm, respectively. The sampling factors for the three filters are approximately $0.504$, $0.617$, and $0.718$ respectively, as indicated by vertical dotted lines in Fig.~\ref{fig:money_plot}. As shown in Fig.~\ref{fig:money_plot}, we obtain a bias of about 0.015 and 0.025 for $F$ and $H$, bands respectively. For the bluest bandpass $J$, all of our measurements made using our set-up were failures, even in the control branch. Extrapolating the data points to higher sampling factors gives a bias estimate of about 0.1. These estimates may be approximate, but they are sufficient to demonstrate that the bias exceeds well above the much tighter requirement of $3.2\times 10^{-4}$ set in~\cite{Dore2018}.

These dependencies on the galaxy sizes and sampling ratio, combined with the fact that $|m_1| > |m_2|$, strongly suggest that the source of bias is aliasing. The shear measured using adaptive Gaussian moments from metacalibrated exposures can be biased at a few per cent level when the PSF is undersampled, thereby requiring large calibration corrections that need to be calculated from image simulations. This negates the fundamental idea behind \metacal\ of calibrating shear measurements from the data itself without requiring any (or at least large) corrections from image simulations. As explained in Section~\ref{sec:introduction}, the size dependence of aliasing bias introduces a sensitivity to galaxy populations that is otherwise benign for \metacal\ when the PSF is well sampled. Even if the aliasing bias can be calculated accurately with sophisticated simulations, it is of practical inconvenience for $m_1$ and $m_2$ to have very different values, since the multiplicative bias can no longer be treated as a scalar as they are done in cosmological parameter inference studies~\citep[e.g.][]{Hildebrandt2020}. It is therefore expected of \metacal\ to provide shear estimates that are sufficiently unbiased to begin with.  In the following section, we explore some strategies to mitigate the pixelization bias internally without requiring external simulations.

\section{Mitigating aliasing bias}
\label{sec:mitigation}
While the weight function $W({\mathbfit{x}})$ in the definition of moments in equation~(\ref{eq:quad_moments}) is arbitrary in principle, a careful choice of the weight function has to be made in practice. We showed in the previous section that shapes of best-fitting Gaussians to the galaxies, as well as the ellipticities computed from moments using those Gaussians as weight functions show significant aliasing biases. Thus, the choice of $W(\mathbfit{x})$ that maximizes the SNR of the shear estimate is unfortunately not `optimal' due to the bias it incurs when combined with \metacal. We emphasize that this is not because of a mismatch between the light profiles, and can therefore be expected for all model-fitting methods, which share similar underlying principles ~\citep{Simon2017}. In this section, we explore if moment-based shapes estimated with a different choice of Gaussian parameters could lead to unbiased shear, although with a slightly lower SNR.  

\subsection{Non-adaptive weight functions}
\label{sec:nonadaptive}
We argue in Appendix~\ref{app:quad_moments} that moments calculated with wide weight functions should be less susceptible to aliasing, as long as the sampling factor is greater than $0.5$. We restrict ourselves to the case of circular weight functions only, as elliptical weight functions respond strongly to the applied shear whereas the size of the circular weight function is rather robust to the applied shear. The only free parameter for the circular weight function is its width, which we vary. We investigate three different ways of employing a large weight function in practice:
\begin{enumerate}
    \item use a constant weight function, that is sufficiently sampled, for all galaxies (used by~\citealt{Hoekstra2021});
    \item scale the width of the adaptive weight function by a constant;
    \item set a minimum threshold for the width of the weight function.
\end{enumerate}

We emphasize that these choices can be made using the available data themselves, without the need for external simulations. For the remainder of the paper, we focus only on the \ksb\ shear measurement algorithm as the other PSF-correction methods discussed in Section~\ref{sec:psfcorr} do not have the freedom to choose the weight function.

\subsubsection{Constant weight function}
\begin{figure}
    \centering
    \includegraphics[width=\columnwidth]{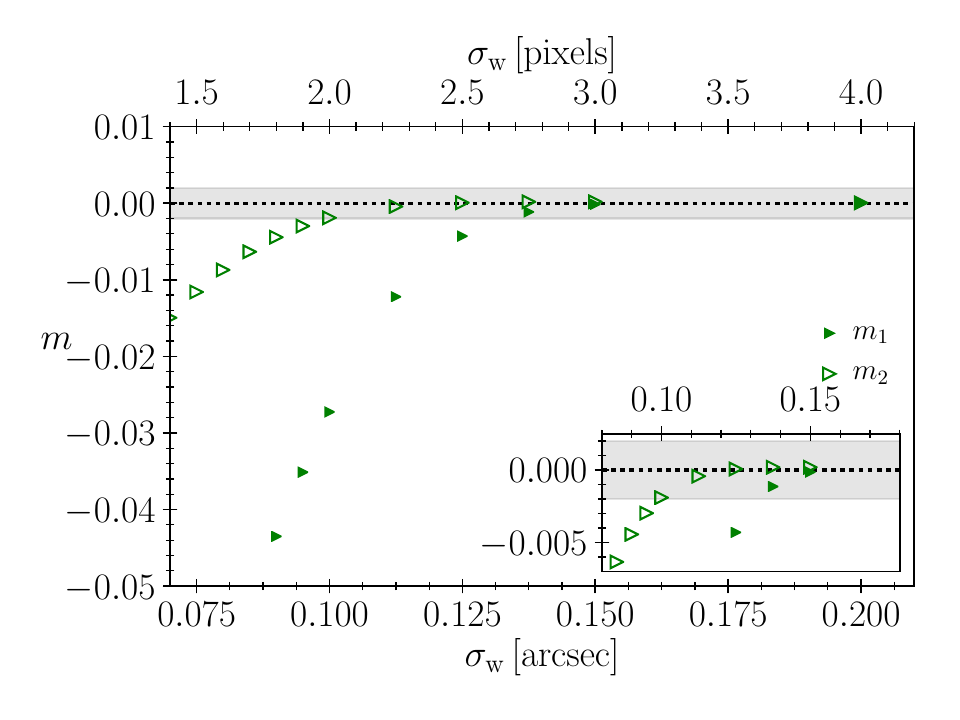}
    \caption{Residual shear multiplicative bias after \metacal\ as a function of $\sigma_{\rm w}$, the width of the circular Gaussian weight function used to compute moments from interleaved images with a pixel scale of $0\farcs{05}$ pixel$^{-1}$. $m_1$ is more adversely affected than $m_2$ due to aliasing for small $\sigma_{\rm w}$. The shaded region represents that \euclid\ requirement on $m$ to be within $\pm 2\times10^{-3}$. The inset focuses on the region of the figure where the bias approaches zero. The horizontal axis of the inset is in units of arcsec. }
    \label{fig:ksbw_size}
\end{figure}
We begin with the simplest scheme of using wide weight functions: a circular Gaussian with a constant width for all galaxies in our sample. This has the advantage of avoiding calculating the adaptive width for each galaxy and is computationally less expensive. We employ the same circular Gaussian function of width $\sigma_{\rm w}$ for all galaxies in the sample, and for all artificially sheared images of them, by specifying the {\texttt{ksb\_sig\_weight}} keyword in \galsim. Figure~\ref{fig:ksbw_size} shows how the bias varies as we change $\sigma_{\rm w}$. A choice of small $\sigma_{\rm w}$ leads to a residual bias in both $m_1$ and $m_2$ components, with $m_1$ being larger in amplitude. A non-negligible level of bias is found in $m_2$ for $\sigma_{\rm w} \lesssim 0\farcs{11}$. The non-zero $m_2$ is because of accentuated aliasing bias even in the $g_2$ shear component.

We find that $\sigma_{\rm w}$ has to be at least $0\farcs{15}$ (for our choice of PSF) for the aliasing bias to be within the allowed uncertainty. For our simulated galaxies, $\sigma_{\rm w} = 0\farcs{15}\  (\sigma_{\rm w} = 0\farcs{2})$ is larger than their adaptive sizes for more than $62\ (38)$ per cent of the galaxies.~\cite{Hoekstra2021} use $\sigma_{\rm w} = 0\farcs{2}$ in their \metacal\ set-up (without any PSF-correction) and find no significant aliasing bias with more complex simulations.

\subsubsection{Constant scaling}
\label{sec:constant_scaling}
\begin{figure}
\centering
\includegraphics[width=\columnwidth]{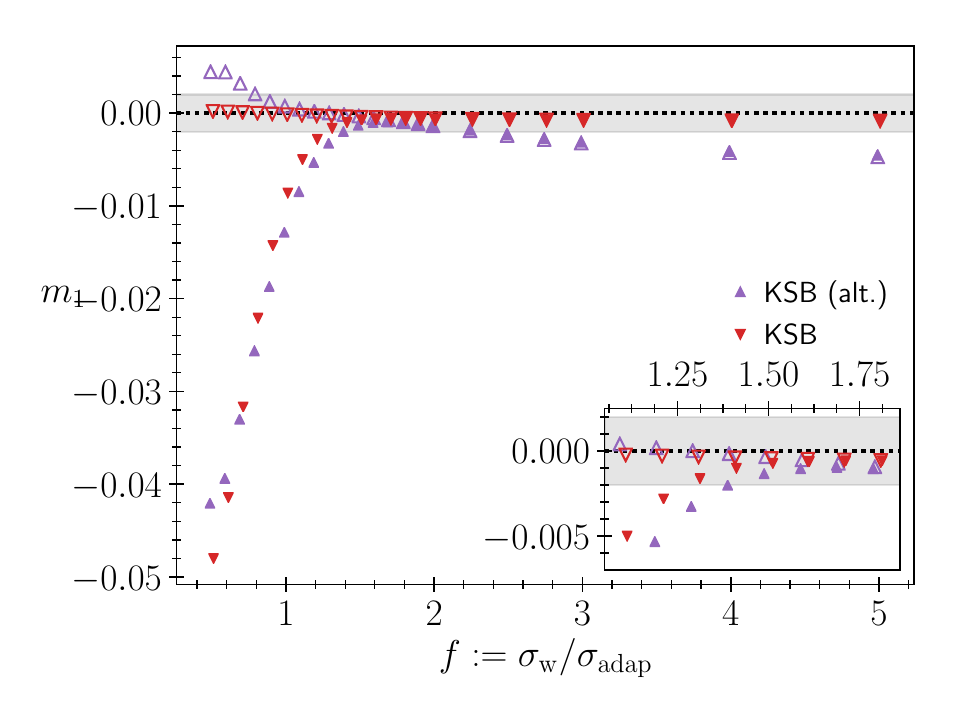}
\caption{Residual shear multiplicative bias post \metacal\ as a function of the multiplicative factor $f$ to scale the size of adaptive weight function. For the points labelled `KSB' (red, downward pointing), the adaptive weight function is measured independently for each metacalibrated exposure, whereas for the points labelled `KSB (alt.)' (purple, upward pointing), the adaptive weight is calculated from the interleaved image composed of exposures not subject to \metacal. The open points denote the bias for the control branch, whereas the filled points denote the \metacal\ branch. The steep behaviour at $f=1$ indicates that adaptive moments are not robust in the presence of aliasing. The shaded region represents the \euclid\ requirement on $m_1$ to be within $\pm 2\times 10^{-3}$. The inset focuses on the region of the figure where the bias approaches zero.}
\label{fig:ksb_size}
\end{figure}

An alternative approach is to scale the width of the adaptive weight function by a constant factor for each galaxy, i.e., $\sigma_{\rm w} = f\sigma_{\rm adap}$. We do this using the {\texttt{ksb\_sig\_factor}} keyword in \galsim. Figure~\ref{fig:ksb_size} shows how the bias changes as a function of $f$. Interestingly, we notice that the bias changes steeply around $f=1$ and approaches $|m_1| < 2\times 10^{-3}$ at around 1.3 and flattens. The smallest bias ($m_1 \approx -6.6\times 10^{-4}$) was achieved for $f = 1.7$ and the bias increases ever so slightly there onwards. The latter trend is true for the control branch as well, indicating that using a very large weight function causes intrinsic bias (perhaps due to truncation) rather than aliasing bias.

Since the adaptive weights themselves are biased, we investigate if the bias could be further reduced by not recomputing the weight function for the sheared versions of the galaxy. For a given galaxy image, we use the adaptive (circular) moments from the interleaved image to give us a good initial estimate for the width of the weight function. We re-measure galaxy shapes with the \ksb\ method, but now with the same weight function (per galaxy). We find that the estimator is biased more compared to the weight function inferred after the \metacal\ procedure. This is understandable as the galaxy is re-convolved with a larger PSF post \metacal\ and therefore the former results in a slightly wider weight function compared to the one obtained from the interleaved image, and hence has smaller bias.

\subsubsection{Minimum thresholding} 
\label{sec:minimum_thresholding}
\begin{figure}
    \centering
    \includegraphics[width=\columnwidth]{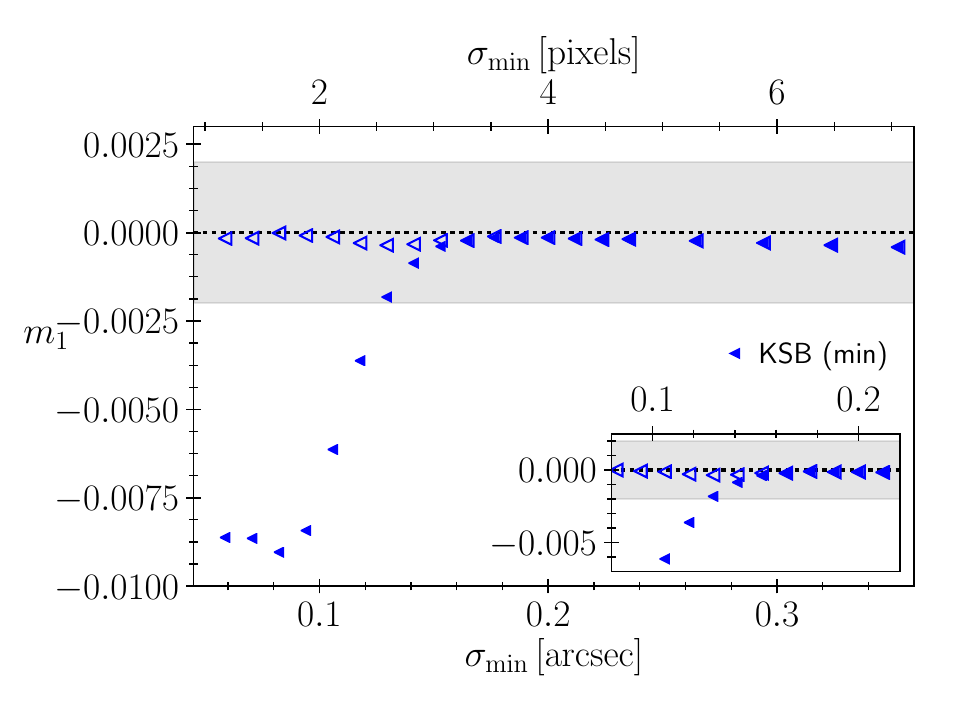}
    \caption{Residual shear multiplicative bias post \metacal\ as a function of $\sigma_{\min}$, the minimum width of the circular Gaussian weight function used to measure moments from the interleaved image with pixel scale $0\farcs{05}$ pixel$^{-1}$. The greater quantity between adaptive size and the minimum threshold size is used for moment calculations. The open points denote the bias for the control branch, which are all within requirements, whereas the filled points denote the \metacal\ branch. The shaded region represents that \euclid\ requirement on $m_1$ to be within $\pm 2\times 10^{-3}$. The inset focuses on the region of the figure where the bias approaches zero. The horizontal axis of the inset is in units of arcsec.}
    \label{fig:ksbm_size}
\end{figure}

Since the use of a constant weight function leads to loss in signal-to-noise for large galaxies that do not contribute to aliasing bias, we devise a hybrid scheme that treats small galaxies differently from the large ones. We propose that the width of the weight function be $\sigma_{\rm w} = \max( \sigma_{\rm \min}, \sigma_{\text{adap}})$, where $\sigma_{\text{adap}}$ is the width of the circular adaptive weight function and $\sigma_{\rm \min}$ is a free parameter to be optimized. We use the \ksb\ algorithm to estimate shear from the ensemble of galaxies using circular Gaussian weight functions with several different values of $\sigma_{\rm \min}$. Figure~\ref{fig:ksbm_size} indicates the multiplicative bias post-\metacal\ as a function of $\sigma_{\rm \min}$. Beyond $\sigma_{\rm \min}$ of $0\farcs{15}$, the bias appears to saturate to a small value that is consistent with zero at the $10^{-4}$ level, and is in excellent agreement with the control branch as well. Comparing with the results in Fig.~\ref{fig:ksbw_size}, it appears that by avoiding Gaussian weight functions smaller than $0\farcs{15}$, we can bring aliasing bias consistent with zero. The exact threshold certainly depends on the size of the PSF (and to some extent the galaxy size distribution), which we analyse next.

\begin{figure}
    \centering
    \includegraphics[width=\columnwidth]{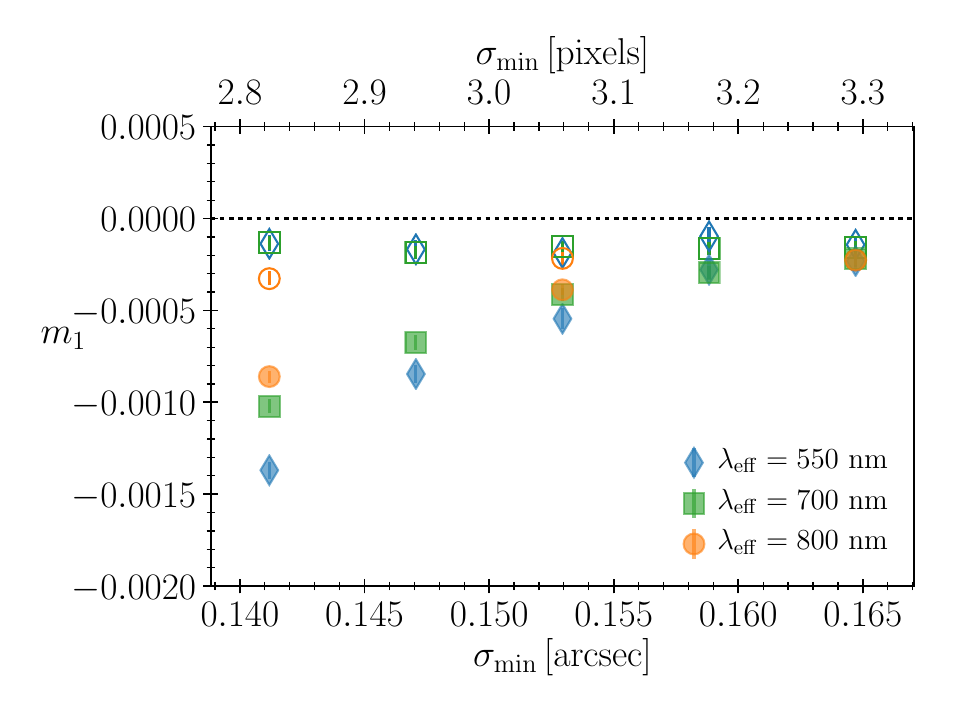}
    \caption{Same as Fig.~\ref{fig:ksbm_size} but for monochromatic PSFs corresponding to shorter wavelengths.}
    \label{fig:ksbm_size_psfs}
\end{figure}

\subsection{Robustness to galaxy colour}
Since the diffraction-limited PSF depends on the galaxy colour, the optimal choices for $\sigma_{\rm w}$ determined above to mitigate the aliasing bias have to be insensitive over a broad range of PSFs in order for these mitigation strategies to be more generally applicable. In particular, they should hold well for PSFs smaller than the one we have considered. In reality, the PSF is the integral of monochromatic PSFs weighted by the SED of the galaxy and filter throughput. The band-limit is then set by the bluest wavelength determined by the SED of the galaxy and VIS passband. We repeat the simulations of the same population of galaxies with Airy PSFs with effective wavelengths $\lambda_\text{eff}$ of $700$\,nm and $550$\,nm.  For these shorter wavelength PSFs, we increased the PSF image oversampling from a factor of 5 to 10 to construct a more accurate interpolated PSF model.  We show the results only for the minimum thresholding strategy as we expect it to be the most robust. From Fig.~\ref{fig:ksbm_size_psfs}, we see only a weak dependence on the effective wavelength, with shorter effective wavelengths showing slightly larger biases for a given minimum $\sigma_{\rm w}$ as expected.  We found the PSF-dependence to be negligible with other strategies as well. Note that $550$\,nm is the shortest wavelength that will be covered with the \euclid\ VIS instrument, and hence this case represents a scenario worse than any realistic worst possible case. It is also reassuring that our mitigation strategy works for strongly undersampled PSFs ($Q \approx 0.47$ for $\lambda_\text{eff} = 550 $\,nm).

One additional complication that we only consider briefly in Appendix~\ref{app:morphology} is that of realistic galaxy morphologies. Star-forming galaxies that suffer more from aliasing due to their blue colour also exhibit features such as knots and spiral arms that boost the power in the high-frequency modes, thus aggravating the impact of  aliasing. We show in Appendix~\ref{app:morphology} that our mitigating strategies are effective for achromatic galaxies. We expect our mitigation strategies to be effective even in the chromatic case, but this is left for future work.

While the true profiles of the PSFs may be different from the ones considered in the study, they are nevertheless band-limited. This is true in the presence of optical aberrations. Optical imperfections such as guiding errors or internal reflections from dichroic coating result in imperfect knowledge of the PSF, but do not affect the band-limit. Out-of-band wavelengths shorter than $550$\,nm increase the band-limit by a small amount. On the other hand, non-ideal instrumental effects such as nonlinear and non-uniform pixel response in real-space that are not perfectly corrected for can mix the various frequency modes thereby introducing a small amount of power at arbitrarily high spatial frequencies. Furthermore, the interpolated image from a finite-size stamp of the PSF itself is not strictly band-limited. However, since we reconvolve the image with a perfectly band-limited PSF, these are not major concerns.

Due to the wavelength dependence, the PSF could vary spatially within a galaxy due to the spatial variation of the SED across the galaxy, referred to as colour gradients~\citep{Voigt2012}. This introduces shear biases larger than what Stage IV experiments can tolerate when a spatially averaged PSF is used to deconvolve the galaxy image. Recently,~\cite{Er2018} showed that \metacal, by means of deconvolving with a wavelength-averaged PSF, cannot fully eliminate biases due to colour gradients, which alone happens to exceed this requirement~\citep{Semboloni2013}. One way around this limitation, at least for moment-based methods, is to use large weight functions that can trade colour gradient bias for noise bias, which can then be removed by \metacal\ in principle~\citep{Semboloni2013,Er2018,Kamath2020}. It is intriguing (and comforting) to see that both aliasing bias and colour gradient biases can be minimised by the same procedure of using a wider weight function, although for different reasons. The low sensitivity to PSFs in Fig.~\ref{fig:ksbm_size_psfs} indicates that not only are our strategies independent of the colour of the galaxy, but should be robust to any colour gradients in the galaxy as well.

\subsection{Robustness to pixel noise}
\label{sec:noisy}
While noiseless image simulations are sufficient to show that \metacal\ suffers from aliasing bias when adaptive moments are used, the mitigation strategies discussed in this section need to be validated on more realistic images with pixel noise in order to demonstrate their usefulness in practice. In particular, the use of wide weight functions could lead to noisier ellipticities and responsitivities, and the shear estimate given in equation~(\ref{eq:shear_estimator}) may no longer be sufficiently unbiased. We now demonstrate using noisy images of isolated galaxies that there is a range of widths for the weight function where the total shear bias is within the requirements. Sections 6 and 7 of our companion paper~\citep{Hoekstra2021} also validate the use of $\sigma_{\rm w} = 0\farcs{2}$ using images that simulate the full field of view.

For simplicity, we assume that the noise is purely additive. We model the noise as Gaussian white noise. The SNR is a fairly complicated function of the galaxy magnitude, size, and ellipticity in addition to the pixel noise. Here, we define SNR $\nu$ as
\begin{equation}
    \nu := \frac{1}{N_\text{side}\sigma_\text{pix}}\sqrt{ \sum\limits_\mathbfit{x} I^2(\mathbfit{x}) } \, ,
\end{equation}
where $\sigma_\text{pix}$ is the root mean square (rms) of the noise per pixel and $I(\mathbfit{x})$ stand for the pixel values of an $N_\text{side} \times N_\text{side}$ image. This definition assumes the use of a matched weight function, which does not hold true as we vary the weight function. Here, we use $\nu$ a proxy for the quality of the galaxy image. For an image consistent with noise, $\nu \approx 1$. We vary the noise rms for each galaxy (using \texttt{galsim.Image.addNoiseSNR} function) so that each galaxy in the sample has a given value of $\nu$.

Since our bias estimate with the small sample is too noisy in the presence of pixel noise, we made the following changes to our simulation set-up described in Section~\ref{sec:method}. We use an input shear of larger magnitude ($g=-0.05-{\rm i}0.05$) in order to reduce the uncertainty in our estimate of the multiplicative bias. Based on the values in Table~\ref{tab:bias_values}, we expect the higher-order shear contribution to increase the multiplicative bias by around 0.001, much smaller than the aliasing bias. We also augment our sample by rendering each galaxy pair thrice with different noise realizations, which gives us small enough uncertainties that are comparable to the requirements of the Stage IV surveys. We continue to use a stimulus shear of magnitude 0.01. To correct for noise correlations introduced during this shearing process, we follow~\cite{Sheldon2017} in applying a noise image sheared in opposite direction.

As in the noiseless case, we take a simple mean of ellipticities and responsitivities, and hence, each galaxy contributes equally to the shear estimate. The justification is that we do not mix galaxies of different SNRs, because the galaxy weights within a sample are expected to be roughly the same\footnote{This is not strictly true, as $\nu$ is detection significance, and galaxies have different shape errors based on their intrinsic properties.}. Having identical weights has the added advantage that the rotated pairs have the same weights as well, thereby ensuring effective shape noise cancellation. 

The uncertainties in the estimates of the shear, and hence in $m_1$, is calculated resampling the pairs of galaxies that are rotated by $90^\circ$, which also ensures shape noise cancellation within each bootstrap realization. We are unable to use the approach of~\cite{Pujol2019} or the more common linear regression approach since we simulated images with only one value of true shear. The bootstrapped shear estimates are Gaussian distributed to a very good approximation. The $1\sigma$ error bars correspond to the 16 - 84 percentile confidence interval over the bootstrap samples. 

\begin{figure}
    \centering
    \includegraphics[width=\columnwidth]{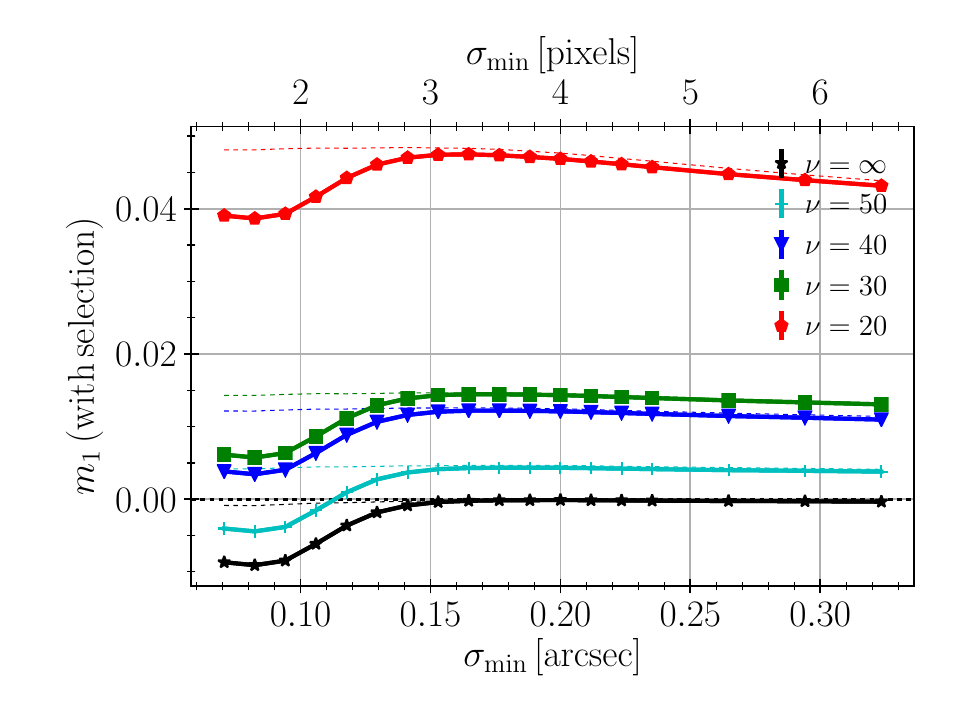}
    \caption{Multiplicative bias due to selection from noisy images applied to noiseless measurements. The selection bias increases in the positive direction as the noise increases, fairly independent of minimum size of the weight function, $\sigma_{\rm \min}$. The uncertainty in selection bias is smaller than the size of markers. The dashed line shows the selection bias on $m_2$. The differences seen at low values of $\sigma_{\rm \min}$ are due to aliasing.}
    \label{fig:ksbm_size_SB}
\end{figure}

In the presence of the noise, the shape measurement failures increased from none to up to $5$ per cent, thereby somewhat changing the selection function. The change in the bias then has (at least) two new components - a selection bias component and a noise bias component.  We also found that the measurement failures had weak dependence on galaxy size. We quantify the change in bias due to a change in the selection function by selecting those galaxy pairs from the noiseless simulations that have valid shape measurements when the SNR is a given value $\nu$. We see from Fig.~\ref{fig:ksbm_size_SB} that as the level of noise increases, imposing the selection on noiseless simulations increases the bias in the positive direction.

The presence of a large residual selection bias is not an inherent limitation of \metacal. ~\cite{Sheldon2017} prescribe how to account for this selection bias, which we have omitted from our analysis for simplicity. A more detailed analysis of selection bias, including those arising from object detection, is presented in~\cite{Hoekstra2021}.

In Table~\ref{tab:ksb_size_snr}, we report the shear multiplicative biases with various weighting schemes from images with finite SNRs. We find that the hybrid scheme (Section~\ref{sec:minimum_thresholding}) is most robust in the presence of pixel noise, with uncertainties much smaller compared to scaling the weight function (Section~\ref{sec:constant_scaling}). We do not report the values from using constant weight functions as the estimates are not robust, and exhibit significant outliers. The lack of robustness is mainly due to the PSF-correction steps within the \ksb\ algorithm, and~\cite{Hoekstra2021} find the use of constant weight functions robust in the absence of explicit PSF-correction.

\begin{figure*}
\centering
\includegraphics[width=\textwidth]{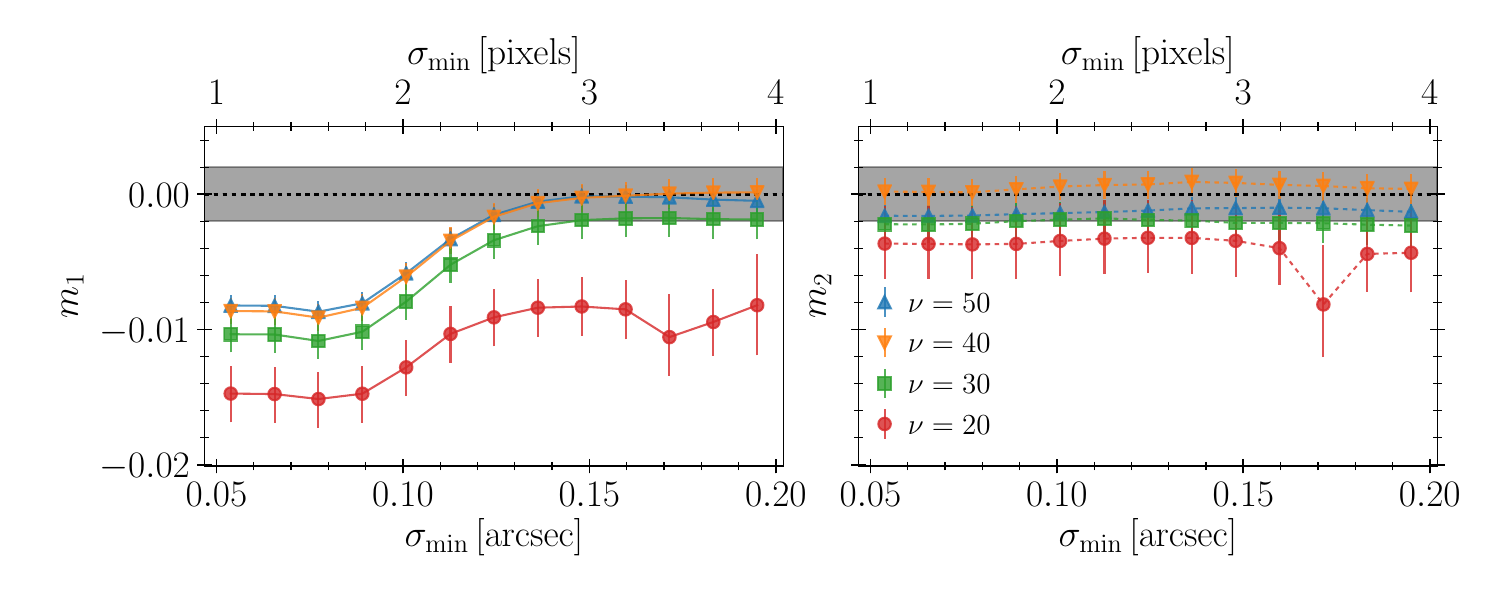}
\caption{Plot of change in multiplicative bias with change in $\sigma_{\rm \min}$ parameter for different values of the SNR $\nu$. The shaded regions represent the tolerance of $\pm 2\times10^{-3}$. The change in bias on noisy images closely follows the trend in the noiseless images for $\sigma_{\rm \min} \lesssim 0\farcs{15}$, suggesting the validity of the mitigation strategy even in the presence of pixel noise.}
\label{fig:size_SNR_KSB}
\end{figure*}

\begin{table}
    \centering
    \caption{Multiplicative biases in the presence of noise for a selected set of schemes that mitigated aliasing bias sufficiently in noiseless images. The multiplicative biases are estimated using equation~(\ref{eq:one_point_estimator}).}
    \label{tab:ksb_size_snr}
    \begin{tabular}{lrrr}
    \hline
    KSB weight & $\nu$ & $m_1$ [$\times 10^2$] & $m_2$ [$\times 10^2$] \\
    \hline
        $\sigma_{\rm w} = \sigma_\text{adap}$ & 50 & $ -0.82 \pm 0.08$ & $ -0.16 \pm 0.08$\\
        $\sigma_{\rm w} = 1.4\sigma_\text{adap}$ & 50 & $ -0.22 \pm 0.09$ & $-0.38 \pm 0. 09$ \\
        $\sigma_{\rm \min} = 0\farcs{15}$ & 50 & $-0.01 \pm 0.08$ & $-0.1 \pm 0.08$\\
        \hline
        $\sigma_{\rm w} = \sigma_\text{adap}$ & 40 & $-0.86 \pm 0.10$ & $0.02 \pm 0.10$ \\
        $\sigma_{\rm w} = 1.4\sigma_\text{adap}$ & 40 & $-0.13 \pm 0.11$ & $-0.08 \pm 0.12$ \\
        $\sigma_{\min} = 0\farcs{15}$ & 40 & $-0.03 \pm 0.10$ & $0.08 \pm 0.10$ \\
        \hline
        $\sigma_{\rm w} = \sigma_\text{adap}$ & 30 & $ -1.03 \pm 0.13$ & $ -0.22 \pm 0.14$\\
        $\sigma_{\rm w} = 1.4\sigma_\text{adap}$ & 30 & $ -0.45 \pm 0.16$ & $-0.35 \pm 0.15$ \\
        $\sigma_{\rm \min} = 0\farcs{15}$ & 30 & $-0.19 \pm 0.14$ & $-0.21 \pm 0.14$\\
        \hline
        $\sigma_{\rm w} = \sigma_\text{adap}$ & 20 & $ -1.47 \pm 0.21$ & $ -0.36 \pm 0.27$\\
        $\sigma_{\rm w} = 1.4\sigma_\text{adap}$ & 20 & $ -2.41 \pm 0.86$ & $-5.20 \pm 2.02$ \\
        $\sigma_{\rm \min} = 0\farcs{15}$ & 20 & $-0.83 \pm 0.21$ & $-0.34 \pm 0.27$\\
        \hline 
    \end{tabular}
\end{table}

Figure~\ref{fig:size_SNR_KSB} shows the bias in the shear estimated from our \metacal\ set-up with the hybrid scheme. Note that the data points are highly correlated, since the measurements are made from the same set of noisy images. The characteristic shape of aliasing bias is seen in $m_1$ values when $\sigma_{\rm \min} \lesssim 0\farcs{125}$. The signature of aliasing bias is absent for $m_2$, as expected from the noiseless case, and is largely independent of the choice of $\sigma_{\rm \min}$. The values of $m_1$ and $m_2$ are also consistent with each other for $\sigma_{\rm \min} \gtrsim 0\farcs{125}$, and marginally so for the $\nu = 20$ sample. The uncertainty in the $m$ values are smaller than the tolerance for $\nu \ge 30$ galaxy samples. The values of $m_2$ (and $m_1$ for large enough $\sigma_{\rm \min}$) are also consistent with 0 for high SNR ($\nu \gtrsim 30$) samples, only barely consistent with 0 for $\nu = 30$ sample and inconsistent with 0 for the $\nu=20$ sample. 

Though the performance of \ksb\ with \metacal\ falls short of the requirement for the $\nu = 20$ sample, the mitigation strategy reduces the bias by about 0.008 for all SNR samples. The residual bias for the low SNR sample may be due to selection effects, correlations in the noise etc. With further work, we believe it should be possible to reduce the bias further for $\nu = 20$, but we leave it outside the scope of this paper.

Despite the simplified nature of the simulations used in this work, Fig.~\ref{fig:size_SNR_KSB} makes evident the presence of aliasing bias and the effectiveness of using a wider weight function in mitigating it. One may suspect that in regions where galaxies are clustered, the ringing effects from one galaxy could affect the shape measurements of neighbouring galaxies, which our simulations do not capture. However, we do not find any evidence of significant aliasing bias in~\cite{Hoekstra2021} who simulate a scene at native pixel scale with multiple galaxies and use a wide weight function for shear measurement. Thus, we conclude that undersampling is not a limiting factor to use \metacal\ for measuring shear using individual exposures from space telescopes if appropriate mitigation measures are taken.

\section{Discussion}
\label{sec:discussion}
The problem of the pixel scale being smaller than the Nyquist scale of the PSF is a fundamental one in estimating the weak lensing signal, as it renders all galaxy images undersampled in principle. In the previous section, we demonstrated a set of related strategies that can sufficiently suppress the effects of aliasing with \metacal. In this section, we layout a few other ways of tackling this problem that may be explored further in the future.

First, we revisit the self-imposed restriction that the \metacal\ procedures are to be performed on individual exposures, and not on the coadded image (see Section~\ref{sec:metacal_procedure}). Our analysis in Fig.~\ref{fig:money_plot} shows that the \metacal\ procedure works, at least in principle, if performed on a coadded image. This is due to a smaller effective pixel scale compared to the native pixel scale. For this study, we considered uniformly dithered exposures with identical PSFs, which are effective at cancelling aliasing effects due to special symmetries. It is not obvious in the more generic case of variable PSFs and random subpixels dither if the coadded image would be free of aliasing.  This is made worse when the exposures have to undergo generic geometric transformations, such as rotation and corrections for field distortions in addition to translations before being projected on to a common grid. A commonly used method to produce a coadded image from the individual HST exposures is the {\sc MultiDrizzle} method~\citep{Drizzle, Koekemoer2003}, which generates a high-resolution image by means of interpolation on the individual (non-Nyquist sampled) exposures, and therefore shapes measured on them with \metacal\ would exhibit some aliasing bias. 

Since the coadded image is a linear combination of the individual exposures, one could construct a corresponding coadded PSF using the same linear combination and use that for deconvolution. Metacalibration has been shown to work on such coadds that are linear combination of individual exposures with PSFs that have smooth spatial-variation at the level expected for the LSST survey~\citep[][Armstrong et al. in preparation]{Sheldon2020}. The caveat however is that, with hundreds of exposures as in LSST, the issue of PSF discontinuity could be overcome by discarding the few exposures (around $3$ per cent of exposures) that contribute to the discontinuity for a given galaxy with very little loss of SNR. Such an approach may not be feasible with Euclid with five (or fewer) exposures per field. But fewer exposures also means that regions with PSF discontinuities due to chip gaps are also much rarer. Additionally, pixels that are masked or affected by cosmic rays also lead to PSF discontinuities. However, with variable PSFs, each PSF image would be undersampled, and hence the coadded PSF image would also be aliased. Alternatively, it is possible to construct a coadded image with a desired target PSF, as in the IMage COMbination algorithm~\citep[{\sc imcom};][]{IMCOM_algo}. This introduces additional complications to the noise properties, which have to be dealt with carefully. Evaluating the performance of \metacal\ on coadded images where the individual PSFs are varying and undersampled is therefore still an open question.

Second, the requirement for the PSF to be Nyquist sampled arises only if we wish to reconstruct the continuous image from the pixel values without any assumption about the galaxy profile. This is different from the requirement of having sufficient knowledge of PSF, which can be obtained from forward modelling of the telescope optics in addition to multiple star images. It is possible to overcome the Nyquist theorem by assuming a moderately realistic model for galaxies, as the model-fitting methods do. Such methods assume a galaxy model with a few parameters and attempt to find the best-fitting model parameters by maximizing the likelihood that the observed image is given by a convolution of the model and the PSF. A similar approach should be possible even when the PSF is undersampled. Given an aliased image after the  \metacal\ procedures, by fitting to each exposure an aliased model convolved with an aliased PSF model, the best-fitting parameters could still be obtained, at least in principle, by likelihood-based methods. Additionally, likelihood-based methods also enjoy adopting an accurate description of correlated noise introduced in the image processing steps. However, a plausible difficulty while trying to implement such an approach is that the continuous image becomes periodic, with a periodicity comparable to pixel width. This could hinder likelihood methods from achieving convergence in the presence of noise. Tight priors on the centroid of the galaxy would be required to prevent diverging solutions. In this paper, we did not use any likelihood-based model-fitting methods and are hence defer testing this hypothesis to the future.

Third, if one were to employ inverse variance weights, or weights that are smooth monotonically increasing functions of galaxy sizes and use equation~(\ref{eq:generalized_shear_estimator}), the amplitude of the aliasing bias may be reduced further. This could help achieve a compromise between the systematic and statistical errors. It may even be possible to conceive a somewhat sophisticated colour-dependent weighting scheme for galaxies so that blue galaxies are downweighted more compared to red galaxies (of the same SNR and size) due to their small PSF size. Such weighting schemes are not without caveats. A colour-dependent weight couples errors in photometric colours (and redshifts) with errors in shear which could be difficult to decouple later on.  Also, in the presence of noise, the estimated size of the galaxies will themselves be noisy and a selection based on these noisy quantities could introduce biases comparable to aliasing bias. Moreover, such weighting schemes might preferentially downweight galaxies at high redshifts, thereby potentially reducing the figure of merit for the dark energy equation of state parameters significantly. A thorough investigation of optimal galaxy weights to minimise the overall bias requires far more realistic image simulations and is beyond the scope of this work.

\section{Summary}
The shear field due to the gravitational lensing of the large-scale structure of the Universe contains rich information about the distribution of dark matter and the rate at which it forms the cosmic web. \Metacal\ is a promising way to calibrate the shear from observed images, without having to rely on external image simulations for calibration. In the case of space-based surveys such as \euclid\ and the \wfirst, this procedure is complicated by the fact that one has to interpolate a possibly undersampled image. Galaxies that are small and barely resolved lead to a significant aliasing bias as a result of this interpolation. Nevertheless, by using moments of those galaxies measured with a weight function wider than their adaptive weight function to estimate shear, it is possible to reduce the amplitude of this bias to a tolerable level, at the mild expense of reducing the SNR.  Thus, assuming that the PSF for the individual exposures are well known, we conclude that \metacal, even when applied to individual exposures, is not limited by undersampling. We understand this conclusion as a result of suppression of high-frequency modes which would otherwise masquerade as low-frequency modes when interpolated. 

Colour-gradient bias, an effect arising due to wavelength dependent PSF and spatial variation of SED within a galaxy, is also mitigated by using similarly wide weight functions for a different reason~\citep{Er2018}. This chromatic effect is not unique to space-based surveys, and is also to relevant to LSST~\citep{Meyers2015}. Thus, sampling is the main fundamental difference between space-based and ground-based telescopes. With the strategies proposed in this paper to mitigate the bias due to sampling, the success of \metacal\ for \euclid\ must follow the success for ground-based surveys as well. Our results therefore encourage the possibility of \metacal\ as an independent shear measurement pipeline, which can validate the lensing shear signal measured using methods that are already existing or under development for \euclid.

\section*{Acknowledgements}
The authors are grateful to Erin Sheldon and the developers of \galsim\ for making their software packages publicly available. The authors thank Patricia Liebing and Lance Miller for helpful conversations about \euclid\ PSFs and Peter Melchior and Erin Sheldon for other technical discussions, and Jason Rhodes and the referee Michael Troxel for carefully reading the manuscript and suggesting several improvements. ER thanks the LEAPS programme\footnote{\url{http://leaps.strw.leidenuniv.nl/}} and its organizers. This work was supported by the Netherlands Organisation for Scientific Research (NWO) through grant 639.043.512, and by the National Science Foundation
under Cooperative Agreement 1258333 managed by the Association of
Universities for Research in Astronomy (AURA), and the Department of Energy
under Contract No. DE-AC02-76SF00515 with the SLAC National Accelerator
Laboratory. Additional funding for Rubin Observatory comes from private
donations, grants to universities, and in-kind support from LSSTC Institutional
Members.

\section*{Data Availability}
The {\sc Python} scripts to reproduce the simulations and analyses may be requested from the corresponding author. The external libraries and data products used in this article are publicly available.





\bibliography{master_ref}

\begin{thebibliography}{}
\makeatletter
\relax
\def\mn@urlcharsother{\let\do\@makeother \do\$\do\&\do\#\do\^\do\_\do\%\do\~}
\def\mn@doi{\begingroup\mn@urlcharsother \@ifnextchar [ {\mn@doi@}
  {\mn@doi@[]}}
\def\mn@doi@[#1]#2{\def\@tempa{#1}\ifx\@tempa\@empty \href
  {http://dx.doi.org/#2} {doi:#2}\else \href {http://dx.doi.org/#2} {#1}\fi
  \endgroup}
\def\mn@eprint#1#2{\mn@eprint@#1:#2::\@nil}
\def\mn@eprint@arXiv#1{\href {http://arxiv.org/abs/#1} {{\tt arXiv:#1}}}
\def\mn@eprint@dblp#1{\href {http://dblp.uni-trier.de/rec/bibtex/#1.xml}
  {dblp:#1}}
\def\mn@eprint@#1:#2:#3:#4\@nil{\def\@tempa {#1}\def\@tempb {#2}\def\@tempc
  {#3}\ifx \@tempc \@empty \let \@tempc \@tempb \let \@tempb \@tempa \fi \ifx
  \@tempb \@empty \def\@tempb {arXiv}\fi \@ifundefined
  {mn@eprint@\@tempb}{\@tempb:\@tempc}{\expandafter \expandafter \csname
  mn@eprint@\@tempb\endcsname \expandafter{\@tempc}}}

\bibitem[\protect\citeauthoryear{{Aihara} et~al.,}{{Aihara}
  et~al.}{2018}]{Aihara2018}
{Aihara} H.,  et~al., 2018, \mn@doi [\pasj] {10.1093/pasj/psx066}, \href
  {https://ui.adsabs.harvard.edu/abs/2018PASJ...70S...4A} {70, S4}

\bibitem[\protect\citeauthoryear{{Akeson} et~al.,}{{Akeson}
  et~al.}{2019}]{Akeson2019}
{Akeson} R.,  et~al., 2019, arXiv e-prints, \href
  {https://ui.adsabs.harvard.edu/abs/2019arXiv190205569A} {p. arXiv:1902.05569}

\bibitem[\protect\citeauthoryear{{Albrecht} et~al.,}{{Albrecht}
  et~al.}{2006}]{Albrecht2006}
{Albrecht} A.,  et~al., 2006, arXiv e-prints, \href
  {https://ui.adsabs.harvard.edu/abs/2006astro.ph..9591A} {pp
  astro--ph/0609591}

\bibitem[\protect\citeauthoryear{{Antilogus}, {Astier}, {Doherty}, {Guyonnet}
  \& {Regnault}}{{Antilogus} et~al.}{2014}]{Antilogus2014}
{Antilogus} P.,  {Astier} P.,  {Doherty} P.,  {Guyonnet} A.,   {Regnault} N.,
  2014, \mn@doi [Journal of Instrumentation] {10.1088/1748-0221/9/03/C03048},
  \href {https://ui.adsabs.harvard.edu/abs/2014JInst...9C3048A} {9, C03048}

\bibitem[\protect\citeauthoryear{{Asgari} et~al.,}{{Asgari}
  et~al.}{2021}]{Asgari2021}
{Asgari} M.,  et~al., 2021, \mn@doi [\aap] {10.1051/0004-6361/202039070}, \href
  {https://ui.adsabs.harvard.edu/abs/2021A&A...645A.104A} {645, A104}

\bibitem[\protect\citeauthoryear{Bartelmann \& Schneider}{Bartelmann \&
  Schneider}{2001}]{Bartelmann2001}
Bartelmann M.,  Schneider P.,  2001, \mn@doi [Physics Reports]
  {10.1016/S0370-1573(00)00082-X}, 340, 291

\bibitem[\protect\citeauthoryear{Bernstein}{Bernstein}{2010}]{Bernstein2010}
Bernstein G.,  2010, \mn@doi [\mnras] {10.1111/j.1365-2966.2010.16883.x}, 406,
  2793

\bibitem[\protect\citeauthoryear{Bernstein \& Jarvis}{Bernstein \&
  Jarvis}{2002}]{BJ02}
Bernstein G.,  Jarvis M.,  2002, \mn@doi [\aj] {10.1086/338085}, 123, 583

\bibitem[\protect\citeauthoryear{{Bosch} et~al.,}{{Bosch}
  et~al.}{2018}]{Bosch2018}
{Bosch} J.,  et~al., 2018, \mn@doi [\pasj] {10.1093/pasj/psx080}, \href
  {https://ui.adsabs.harvard.edu/abs/2018PASJ...70S...5B} {70, S5}

\bibitem[\protect\citeauthoryear{Bridle et~al.,}{Bridle et~al.}{2009}]{GREAT08}
Bridle S.,  et~al., 2009, \mn@doi [Annals of Applied Statistics]
  {10.1214/08-AOAS222}, 3, 6

\bibitem[\protect\citeauthoryear{{Carlsten}, {Strauss}, {Lupton}, {Meyers}  \&
  {Miyazaki}}{{Carlsten} et~al.}{2018}]{Carlsten2018}
{Carlsten} S.~G.,  {Strauss} M.~A.,  {Lupton} R.~H.,  {Meyers} J.~E.,
  {Miyazaki} S.,  2018, \mn@doi [\mnras] {10.1093/mnras/sty1636}, \href
  {https://ui.adsabs.harvard.edu/abs/2018MNRAS.479.1491C} {479, 1491}

\bibitem[\protect\citeauthoryear{{Coulton}, {Armstrong}, {Smith}, {Lupton}  \&
  {Spergel}}{{Coulton} et~al.}{2018}]{Coulton2018}
{Coulton} W.~R.,  {Armstrong} R.,  {Smith} K.~M.,  {Lupton} R.~H.,   {Spergel}
  D.~N.,  2018, \mn@doi [\aj] {10.3847/1538-3881/aac08d}, \href
  {https://ui.adsabs.harvard.edu/abs/2018AJ....155..258C} {155, 258}

\bibitem[\protect\citeauthoryear{{Dawson}, {Schneider}, {Tyson}  \&
  {Jee}}{{Dawson} et~al.}{2016}]{Dawson2016}
{Dawson} W.~A.,  {Schneider} M.~D.,  {Tyson} J.~A.,   {Jee} M.~J.,  2016,
  \mn@doi [\apj] {10.3847/0004-637X/816/1/11}, \href
  {https://ui.adsabs.harvard.edu/abs/2016ApJ...816...11D} {816, 11}

\bibitem[\protect\citeauthoryear{{Dor{\'e}} et~al.,}{{Dor{\'e}}
  et~al.}{2018}]{Dore2018}
{Dor{\'e}} O.,  et~al., 2018, arXiv e-prints, \href
  {https://ui.adsabs.harvard.edu/abs/2018arXiv180403628D} {p. arXiv:1804.03628}

\bibitem[\protect\citeauthoryear{{Eckert} et~al.,}{{Eckert}
  et~al.}{2020}]{Eckert2020}
{Eckert} K.,  et~al., 2020, \mn@doi [\mnras] {10.1093/mnras/staa2133}, \href
  {https://ui.adsabs.harvard.edu/abs/2020MNRAS.497.2529E} {497, 2529}

\bibitem[\protect\citeauthoryear{Er et~al.,}{Er et~al.}{2018}]{Er2018}
Er X.,  et~al., 2018, \mn@doi [\mnras] {10.1093/mnras/sty685}, 476, 5645

\bibitem[\protect\citeauthoryear{{Euclid Collaboration} et~al.,}{{Euclid
  Collaboration} et~al.}{2019}]{Martinet2019}
{Euclid Collaboration} et~al., 2019, \mn@doi [\aap]
  {10.1051/0004-6361/201935187}, \href
  {https://ui.adsabs.harvard.edu/abs/2019A&A...627A..59E} {627, A59}

\bibitem[\protect\citeauthoryear{{Fenech Conti}, {Herbonnet}, {Hoekstra},
  {Merten}, {Miller}  \& {Viola}}{{Fenech Conti}
  et~al.}{2017}]{FenechConti2017}
{Fenech Conti} I.,  {Herbonnet} R.,  {Hoekstra} H.,  {Merten} J.,  {Miller} L.,
    {Viola} M.,  2017, \mn@doi [\mnras] {10.1093/mnras/stx200}, \href
  {https://ui.adsabs.harvard.edu/abs/2017MNRAS.467.1627F} {467, 1627}

\bibitem[\protect\citeauthoryear{Fruchter \& Hook}{Fruchter \&
  Hook}{2002}]{Drizzle}
Fruchter A.~S.,  Hook R.~N.,  2002, \mn@doi [\pasp] {10.1086/338393}, 114, 144

\bibitem[\protect\citeauthoryear{Gurvich \& Mandelbaum}{Gurvich \&
  Mandelbaum}{2016}]{Gurvich2016}
Gurvich A.,  Mandelbaum R.,  2016, \mn@doi [\mnras] {10.1093/mnras/stw174},
  457, 3522

\bibitem[\protect\citeauthoryear{Heymans et~al.,}{Heymans et~al.}{2006}]{STEP1}
Heymans C.,  et~al., 2006, \mn@doi [\mnras] {10.1111/j.1365-2966.2006.10198.x},
  368, 1323

\bibitem[\protect\citeauthoryear{{Hikage} et~al.,}{{Hikage}
  et~al.}{2019}]{Hikage2019}
{Hikage} C.,  et~al., 2019, \mn@doi [\pasj] {10.1093/pasj/psz010}, \href
  {https://ui.adsabs.harvard.edu/abs/2019PASJ...71...43H} {71, 43}

\bibitem[\protect\citeauthoryear{{Hildebrandt} et~al.,}{{Hildebrandt}
  et~al.}{2020}]{Hildebrandt2020}
{Hildebrandt} H.,  et~al., 2020, \mn@doi [\aap] {10.1051/0004-6361/201834878},
  \href {https://ui.adsabs.harvard.edu/abs/2020A&A...633A..69H} {633, A69}

\bibitem[\protect\citeauthoryear{Hirata \& Seljak}{Hirata \&
  Seljak}{2003}]{HS03}
Hirata C.,  Seljak U.,  2003, \mn@doi [\mnras]
  {10.1046/j.1365-8711.2003.06683.x}, 343, 459

\bibitem[\protect\citeauthoryear{Hoekstra \& Jain}{Hoekstra \&
  Jain}{2008}]{HoekstraJain2008}
Hoekstra H.,  Jain B.,  2008, \mn@doi [Annual Review of Nuclear and Particle
  Science] {10.1146/annurev.nucl.58.110707.171151}, 58, 99

\bibitem[\protect\citeauthoryear{Hoekstra, Franx, Kuijken  \& Squires}{Hoekstra
  et~al.}{1998}]{Hoekstra1998}
Hoekstra H.,  Franx M.,  Kuijken K.,   Squires G.,  1998, \mn@doi [\apj]
  {10.1086/306102}, 504, 636

\bibitem[\protect\citeauthoryear{{Hoekstra}, {Herbonnet}, {Muzzin}, {Babul},
  {Mahdavi}, {Viola}  \& {Cacciato}}{{Hoekstra} et~al.}{2015}]{Hoekstra2015}
{Hoekstra} H.,  {Herbonnet} R.,  {Muzzin} A.,  {Babul} A.,  {Mahdavi} A.,
  {Viola} M.,   {Cacciato} M.,  2015, \mn@doi [\mnras] {10.1093/mnras/stv275},
  \href {https://ui.adsabs.harvard.edu/abs/2015MNRAS.449..685H} {449, 685}

\bibitem[\protect\citeauthoryear{{Hoekstra}, {Viola}  \&
  {Herbonnet}}{{Hoekstra} et~al.}{2017}]{Hoekstra2017}
{Hoekstra} H.,  {Viola} M.,   {Herbonnet} R.,  2017, \mn@doi [\mnras]
  {10.1093/mnras/stx724}, \href
  {https://ui.adsabs.harvard.edu/abs/2017MNRAS.468.3295H} {468, 3295}

\bibitem[\protect\citeauthoryear{{Hoekstra}, {Kannawadi}  \&
  {Kitching}}{{Hoekstra} et~al.}{2021}]{Hoekstra2021}
{Hoekstra} H.,  {Kannawadi} A.,   {Kitching} T.~D.,  2021, \mn@doi [\aap]
  {10.1051/0004-6361/202038998}, \href
  {https://ui.adsabs.harvard.edu/abs/2021A&A...646A.124H} {646, A124}

\bibitem[\protect\citeauthoryear{{Huff} \& {Mandelbaum}}{{Huff} \&
  {Mandelbaum}}{2017}]{Huff2017}
{Huff} E.,  {Mandelbaum} R.,  2017, arXiv e-prints, \href
  {https://ui.adsabs.harvard.edu/abs/2017arXiv170202600H} {p. arXiv:1702.02600}

\bibitem[\protect\citeauthoryear{Israel et~al.,}{Israel
  et~al.}{2015}]{Israel2015}
Israel H.,  et~al., 2015, \mn@doi [MNRAS] {10.1093/mnras/stv1660}, 453, 561

\bibitem[\protect\citeauthoryear{{Ivezi{\'c}} et~al.,}{{Ivezi{\'c}}
  et~al.}{2019}]{Ivezic2019}
{Ivezi{\'c}} {\v{Z}}.,  et~al., 2019, \mn@doi [\apj]
  {10.3847/1538-4357/ab042c}, \href
  {https://ui.adsabs.harvard.edu/abs/2019ApJ...873..111I} {873, 111}

\bibitem[\protect\citeauthoryear{Kacprzak, Zuntz, Rowe, Bridle, Refregier,
  Amara, Voigt  \& Hirsch}{Kacprzak et~al.}{2012}]{Kacprzak2012}
Kacprzak T.,  Zuntz J.,  Rowe B.,  Bridle S.,  Refregier A.,  Amara A.,  Voigt
  L.,   Hirsch M.,  2012, \mn@doi [\mnras] {10.1111/j.1365-2966.2012.21622},
  427, 2711

\bibitem[\protect\citeauthoryear{{Kaiser}}{{Kaiser}}{2000}]{Kaiser2000}
{Kaiser} N.,  2000, \mn@doi [\apj] {10.1086/309041}, \href
  {https://ui.adsabs.harvard.edu/abs/2000ApJ...537..555K} {537, 555}

\bibitem[\protect\citeauthoryear{Kaiser, Squires  \& Broadhurst}{Kaiser
  et~al.}{1995}]{KSB95}
Kaiser N.,  Squires G.,   Broadhurst T.,  1995, \mn@doi [\apj]
  {10.1086/176071}, 449, 449

\bibitem[\protect\citeauthoryear{{Kamath}, {Meyers}, {Burchat}  \& {(LSST Dark
  Energy Science Collaboration}}{{Kamath} et~al.}{2020}]{Kamath2020}
{Kamath} S.,  {Meyers} J.~E.,  {Burchat} P.~R.,   {(LSST Dark Energy Science
  Collaboration} 2020, \mn@doi [\apj] {10.3847/1538-4357/ab54cb}, \href
  {https://ui.adsabs.harvard.edu/abs/2020ApJ...888...23K} {888, 23}

\bibitem[\protect\citeauthoryear{Kannawadi, Mandelbaum  \& Lackner}{Kannawadi
  et~al.}{2015}]{Kannawadi2015}
Kannawadi A.,  Mandelbaum R.,   Lackner C.,  2015, \mn@doi [\mnras]
  {10.1093/mnras/stv520}, 449, 3597

\bibitem[\protect\citeauthoryear{{Kannawadi}, {Shapiro}, {Mandelbaum},
  {Hirata}, {Kruk}  \& {Rhodes}}{{Kannawadi} et~al.}{2016}]{Kannawadi_IPC_PSF}
{Kannawadi} A.,  {Shapiro} C.~A.,  {Mandelbaum} R.,  {Hirata} C.~M.,  {Kruk}
  J.~W.,   {Rhodes} J.~D.,  2016, \mn@doi [\pasp]
  {10.1088/1538-3873/128/967/095001}, 128, 095001

\bibitem[\protect\citeauthoryear{{Kannawadi} et~al.,}{{Kannawadi}
  et~al.}{2019}]{Kannawadi2019}
{Kannawadi} A.,  et~al., 2019, \mn@doi [\aap] {10.1051/0004-6361/201834819},
  \href {https://ui.adsabs.harvard.edu/abs/2019A&A...624A..92K} {624, A92}

\bibitem[\protect\citeauthoryear{Kilbinger}{Kilbinger}{2015}]{Kilbinger15}
Kilbinger M.,  2015, \mn@doi [Reports on Progress in Physics]
  {10.1088/0034-4885/78/8/086901}, 78, 86901

\bibitem[\protect\citeauthoryear{Kitching et~al.,}{Kitching
  et~al.}{2011}]{Kitching2011}
Kitching T.,  et~al., 2011, \mn@doi [Ann. Appl. Stat.] {10.1214/11-AOAS484}, 5,
  2231

\bibitem[\protect\citeauthoryear{{Koekemoer}, {Fruchter}, {Hook}  \&
  {Hack}}{{Koekemoer} et~al.}{2003}]{Koekemoer2003}
{Koekemoer} A.~M.,  {Fruchter} A.~S.,  {Hook} R.~N.,   {Hack} W.,  2003, in HST
  Calibration Workshop : Hubble after the Installation of the ACS and the
  NICMOS Cooling System. p.~337

\bibitem[\protect\citeauthoryear{Koekemoer et~al.,}{Koekemoer
  et~al.}{2007}]{COSMOS_generic}
Koekemoer A.,  et~al., 2007, \mn@doi [\apjs] {10.1086/520086}, 172, 196

\bibitem[\protect\citeauthoryear{{LSST Science Collaboration} et~al.,}{{LSST
  Science Collaboration} et~al.}{2009}]{LSST_Book}
{LSST Science Collaboration} et~al., 2009, arXiv e-prints, \href
  {https://ui.adsabs.harvard.edu/abs/2009arXiv0912.0201L} {p. arXiv:0912.0201}

\bibitem[\protect\citeauthoryear{Lackner \& Gunn}{Lackner \&
  Gunn}{2012}]{Claire_Fits}
Lackner C.,  Gunn J.,  2012, \mn@doi [\mnras]
  {10.1111/j.1365-2966.2012.20450.x}, 421, 2277

\bibitem[\protect\citeauthoryear{Lauer}{Lauer}{1999}]{Lauer99}
Lauer T.~R.,  1999, \mn@doi [\pasp] {10.1086/316319}, 111, 227

\bibitem[\protect\citeauthoryear{{Laureijs} et~al.,}{{Laureijs}
  et~al.}{2011}]{Laureijs2011}
{Laureijs} R.,  et~al., 2011, arXiv e-prints, \href
  {https://ui.adsabs.harvard.edu/abs/2011arXiv1110.3193L} {p. arXiv:1110.3193}

\bibitem[\protect\citeauthoryear{Leauthaud et~al.,}{Leauthaud
  et~al.}{2007}]{COSMOS_Alexie}
Leauthaud A.,  et~al., 2007, \mn@doi [\apjs] {10.1086/516598}, 172, 219

\bibitem[\protect\citeauthoryear{Luppino \& Kaiser}{Luppino \&
  Kaiser}{1997}]{Luppino1997}
Luppino G.~A.,  Kaiser N.,  1997, \mn@doi [\apj] {10.1086/303508}, 475, 20

\bibitem[\protect\citeauthoryear{{Mandelbaum}}{{Mandelbaum}}{2018}]{Mandelbaum2018a}
{Mandelbaum} R.,  2018, \mn@doi [\araa] {10.1146/annurev-astro-081817-051928},
  \href {https://ui.adsabs.harvard.edu/abs/2018ARA&A..56..393M} {56, 393}

\bibitem[\protect\citeauthoryear{{Mandelbaum}, {Hirata}, {Leauthaud}, {Massey}
  \& {Rhodes}}{{Mandelbaum} et~al.}{2012}]{Mandelbaum2012}
{Mandelbaum} R.,  {Hirata} C.~M.,  {Leauthaud} A.,  {Massey} R.~J.,   {Rhodes}
  J.,  2012, \mn@doi [\mnras] {10.1111/j.1365-2966.2011.20138.x}, \href
  {https://ui.adsabs.harvard.edu/abs/2012MNRAS.420.1518M} {420, 1518}

\bibitem[\protect\citeauthoryear{Mandelbaum et~al.,}{Mandelbaum
  et~al.}{2014}]{GREAT3}
Mandelbaum R.,  et~al., 2014, \mn@doi [\apjs] {10.1088/0067-0049/212/1/5}, 212,
  5

\bibitem[\protect\citeauthoryear{Mandelbaum et~al.,}{Mandelbaum
  et~al.}{2015}]{GREAT3_results1}
Mandelbaum R.,  et~al., 2015, \mn@doi [\mnras] {10.1093/mnras/stv781}, 450,
  2963

\bibitem[\protect\citeauthoryear{{Mandelbaum} et~al.,}{{Mandelbaum}
  et~al.}{2018}]{Mandelbaum2018}
{Mandelbaum} R.,  et~al., 2018, \mn@doi [\mnras] {10.1093/mnras/sty2420}, \href
  {https://ui.adsabs.harvard.edu/abs/2018MNRAS.481.3170M} {481, 3170}

\bibitem[\protect\citeauthoryear{Massey, Rowe, Refregier, Bacon  \&
  Berg{\'{e}}}{Massey et~al.}{2007}]{Massey2007}
Massey R.,  Rowe B.,  Refregier A.,  Bacon D.~J.,   Berg{\'{e}} J.,  2007,
  \mn@doi [\mnras] {10.1111/j.1365-2966.2007.12072.x}, 380, 229

\bibitem[\protect\citeauthoryear{Melchior \& Viola}{Melchior \&
  Viola}{2012}]{Melchior2012}
Melchior P.,  Viola M.,  2012, \mn@doi [\mnras]
  {10.1111/j.1365-2966.2012.21381.x}, 424, 2757

\bibitem[\protect\citeauthoryear{{Meyers} \& {Burchat}}{{Meyers} \&
  {Burchat}}{2015}]{Meyers2015}
{Meyers} J.~E.,  {Burchat} P.~R.,  2015, \mn@doi [\apj]
  {10.1088/0004-637X/807/2/182}, \href
  {https://ui.adsabs.harvard.edu/abs/2015ApJ...807..182M} {807, 182}

\bibitem[\protect\citeauthoryear{Miller, Kitching, Heymans, Heavens, Waerbeke
  \& van Waerbeke}{Miller et~al.}{2007}]{Miller2007}
Miller L.,  Kitching T.~D.,  Heymans C.,  Heavens A.~F.,  Waerbeke L.~V.,   van
  Waerbeke L.,  2007, \mn@doi [\mnras] {10.1111/j.1365-2966.2007.12363.x}, 382,
  315

\bibitem[\protect\citeauthoryear{Miller et~al.,}{Miller
  et~al.}{2013}]{Miller2013}
Miller L.,  et~al., 2013, \mn@doi [\mnras] {10.1093/mnras/sts454}, 429, 2858

\bibitem[\protect\citeauthoryear{Plazas}{Plazas}{2020}]{Plazas2020}
Plazas A.~A.,  2020, \mn@doi [Symmetry] {10.3390/sym12040494}, \href
  {https://ui.adsabs.harvard.edu/abs/2020arXiv200306090P} {12, 494}

\bibitem[\protect\citeauthoryear{{Plazas}, {Shapiro}, {Kannawadi},
  {Mandelbaum}, {Rhodes}  \& {Smith}}{{Plazas} et~al.}{2016}]{Plazas2016}
{Plazas} A.~A.,  {Shapiro} C.,  {Kannawadi} A.,  {Mandelbaum} R.,  {Rhodes} J.,
    {Smith} R.,  2016, \mn@doi [\pasp] {10.1088/1538-3873/128/968/104001},
  \href {https://ui.adsabs.harvard.edu/abs/2016PASP..128j4001P} {128, 104001}

\bibitem[\protect\citeauthoryear{{Pujol}, {Kilbinger}, {Sureau}  \&
  {Bobin}}{{Pujol} et~al.}{2019}]{Pujol2019}
{Pujol} A.,  {Kilbinger} M.,  {Sureau} F.,   {Bobin} J.,  2019, \mn@doi [\aap]
  {10.1051/0004-6361/201833740}, \href
  {https://ui.adsabs.harvard.edu/abs/2019A&A...621A...2P} {621, A2}

\bibitem[\protect\citeauthoryear{{Rhodes}, {Leauthaud}, {Stoughton}, {Massey},
  {Dawson}, {Kolbe}  \& {Roe}}{{Rhodes} et~al.}{2010}]{Rhodes2010}
{Rhodes} J.,  {Leauthaud} A.,  {Stoughton} C.,  {Massey} R.,  {Dawson} K.,
  {Kolbe} W.,   {Roe} N.,  2010, \mn@doi [\pasp] {10.1086/651675}, \href
  {https://ui.adsabs.harvard.edu/abs/2010PASP..122..439R} {122, 439}

\bibitem[\protect\citeauthoryear{Rowe, Hirata  \& Rhodes}{Rowe
  et~al.}{2011}]{IMCOM_algo}
Rowe B.,  Hirata C.,   Rhodes J.,  2011, \mn@doi [\apj]
  {10.1088/0004-637X/741/1/46}, 741, 46

\bibitem[\protect\citeauthoryear{Rowe et~al.,}{Rowe et~al.}{2015}]{galsim}
Rowe B.,  et~al., 2015, \mn@doi [Astronomy and Computing]
  {10.1016/j.ascom.2015.02.002}, 10, 121

\bibitem[\protect\citeauthoryear{{Samuroff} et~al.,}{{Samuroff}
  et~al.}{2018}]{Samuroff2018}
{Samuroff} S.,  et~al., 2018, \mn@doi [\mnras] {10.1093/mnras/stx3282}, \href
  {https://ui.adsabs.harvard.edu/abs/2018MNRAS.475.4524S} {475, 4524}

\bibitem[\protect\citeauthoryear{{S{\'a}nchez} et~al.,}{{S{\'a}nchez}
  et~al.}{2020}]{Sanchez2020}
{S{\'a}nchez} J.,  et~al., 2020, \mn@doi [\mnras] {10.1093/mnras/staa1957},
  \href {https://ui.adsabs.harvard.edu/abs/2020MNRAS.497..210S} {497, 210}

\bibitem[\protect\citeauthoryear{{Schneider} \& {Seitz}}{{Schneider} \&
  {Seitz}}{1995}]{Schneider1995}
{Schneider} P.,  {Seitz} C.,  1995, \aap, \href
  {https://ui.adsabs.harvard.edu/abs/1995A&A...294..411S} {294, 411}

\bibitem[\protect\citeauthoryear{Scoville et~al.,}{Scoville
  et~al.}{2007}]{COSMOS_overview}
Scoville N.,  et~al., 2007, \mn@doi [\apjs] {10.1086/516585}, 172, 1

\bibitem[\protect\citeauthoryear{{Seitz} \& {Schneider}}{{Seitz} \&
  {Schneider}}{1997}]{Seitz1997}
{Seitz} C.,  {Schneider} P.,  1997, \aap, \href
  {https://ui.adsabs.harvard.edu/abs/1997A&A...318..687S} {318, 687}

\bibitem[\protect\citeauthoryear{Semboloni et~al.,}{Semboloni
  et~al.}{2013}]{Semboloni2013}
Semboloni E.,  et~al., 2013, \mn@doi [\mnras] {10.1093/mnras/stt602}, 432, 2385

\bibitem[\protect\citeauthoryear{{S\'{e}rsic}}{{S\'{e}rsic}}{1968}]{Sersic1968}
{S\'{e}rsic} J.~L.,  1968, {Atlas de Galaxias Australes}

\bibitem[\protect\citeauthoryear{Shapiro, Rowe, Goodsall, Hirata, Fucik,
  Rhodes, Seshadri  \& Smith}{Shapiro et~al.}{2013}]{IMCOM_WLsystematics}
Shapiro C.,  Rowe B.,  Goodsall T.,  Hirata C.,  Fucik J.,  Rhodes J.,
  Seshadri S.,   Smith R.,  2013, \mn@doi [\pasp] {10.1086/674415}, 125, 1496

\bibitem[\protect\citeauthoryear{{Sheldon}}{{Sheldon}}{2015}]{Sheldon2015}
{Sheldon} E.,  2015, {NGMIX: Gaussian mixture models for 2D images} (\mn@eprint
  {ascl} {1508.008})

\bibitem[\protect\citeauthoryear{{Sheldon} \& {Huff}}{{Sheldon} \&
  {Huff}}{2017}]{Sheldon2017}
{Sheldon} E.~S.,  {Huff} E.~M.,  2017, \mn@doi [\apj]
  {10.3847/1538-4357/aa704b}, \href
  {https://ui.adsabs.harvard.edu/abs/2017ApJ...841...24S} {841, 24}

\bibitem[\protect\citeauthoryear{{Sheldon}, {Becker}, {MacCrann}  \&
  {Jarvis}}{{Sheldon} et~al.}{2020}]{Sheldon2020}
{Sheldon} E.~S.,  {Becker} M.~R.,  {MacCrann} N.,   {Jarvis} M.,  2020, \mn@doi
  [\apj] {10.3847/1538-4357/abb595}, \href
  {https://ui.adsabs.harvard.edu/abs/2020ApJ...902..138S} {902, 138}

\bibitem[\protect\citeauthoryear{{Simon} \& {Schneider}}{{Simon} \&
  {Schneider}}{2017}]{Simon2017}
{Simon} P.,  {Schneider} P.,  2017, \mn@doi [\aap]
  {10.1051/0004-6361/201629591}, \href
  {https://ui.adsabs.harvard.edu/abs/2017A&A...604A.109S} {604, A109}

\bibitem[\protect\citeauthoryear{Spergel et~al.,}{Spergel
  et~al.}{2015}]{spergel2015wide}
Spergel D.,  et~al., 2015, preprint (\mn@eprint {arXiv} {1503.03757})

\bibitem[\protect\citeauthoryear{{The Dark Energy Survey Collaboration}}{{The
  Dark Energy Survey Collaboration}}{2005}]{DES}
{The Dark Energy Survey Collaboration} 2005, arXiv e-prints, \href
  {https://ui.adsabs.harvard.edu/abs/2005astro.ph.10346T} {pp
  astro--ph/0510346}

\bibitem[\protect\citeauthoryear{{Troxel} et~al.,}{{Troxel}
  et~al.}{2018}]{Troxel2018}
{Troxel} M.~A.,  et~al., 2018, \mn@doi [\prd] {10.1103/PhysRevD.98.043528},
  \href {https://ui.adsabs.harvard.edu/abs/2018PhRvD..98d3528T} {98, 043528}

\bibitem[\protect\citeauthoryear{{Troxel} et~al.,}{{Troxel}
  et~al.}{2021}]{Troxel2021}
{Troxel} M.~A.,  et~al., 2021, \mn@doi [\mnras] {10.1093/mnras/staa3658}, \href
  {https://ui.adsabs.harvard.edu/abs/2021MNRAS.501.2044T} {501, 2044}

\bibitem[\protect\citeauthoryear{Voigt \& Bridle}{Voigt \&
  Bridle}{2010}]{Voigt2010}
Voigt L.,  Bridle S.,  2010, \mn@doi [\mnras]
  {10.1111/j.1365-2966.2010.16300.x}, 404, 458

\bibitem[\protect\citeauthoryear{Voigt, Bridle, Amara, Cropper, Kitching,
  Massey, Rhodes  \& Schrabback}{Voigt et~al.}{2012}]{Voigt2012}
Voigt L.,  Bridle S.,  Amara A.,  Cropper M.,  Kitching T.,  Massey R.,  Rhodes
  J.,   Schrabback T.,  2012, \mn@doi [\mnras]
  {10.1111/j.1365-2966.2011.20395.x}, 421, 1385

\bibitem[\protect\citeauthoryear{Zhang}{Zhang}{2008}]{Zhang2008}
Zhang J.,  2008, \mn@doi [\mnras] {10.1111/j.1365-2966.2007.12585.x}, 383, 113

\bibitem[\protect\citeauthoryear{Zuntz, Kacprzak, Voigt, Hirsch, Rowe  \&
  Bridle}{Zuntz et~al.}{2013}]{Zuntz13}
Zuntz J.,  Kacprzak T.,  Voigt L.,  Hirsch M.,  Rowe B.,   Bridle S.,  2013,
  \mn@doi [\mnras] {10.1093/mnras/stt1125}, 434, 1604

\bibitem[\protect\citeauthoryear{{Zuntz} et~al.,}{{Zuntz}
  et~al.}{2018}]{Zuntz2018}
{Zuntz} J.,  et~al., 2018, \mn@doi [\mnras] {10.1093/mnras/sty2219}, \href
  {https://ui.adsabs.harvard.edu/abs/2018MNRAS.481.1149Z} {481, 1149}

\bibitem[\protect\citeauthoryear{{de Jong}, {Verdoes Kleijn}, {Kuijken}  \&
  {Valentijn}}{{de Jong} et~al.}{2013}]{deJong2013}
{de Jong} J. T.~A.,  {Verdoes Kleijn} G.~A.,  {Kuijken} K.~H.,   {Valentijn}
  E.~A.,  2013, \mn@doi [Experimental Astronomy] {10.1007/s10686-012-9306-1},
  \href {https://ui.adsabs.harvard.edu/abs/2013ExA....35...25D} {35, 25}

\makeatother
\end{thebibliography}



\appendix

\section{Impact due to the choice of the interpolation kernel}
\label{app:interpolation}

In principle, when a discretized signal is convolved with a \texttt{sinc} kernel, it reproduces the original signal in the continuous domain, provided that the signal was Nyquist-sampled. However, such a perfect interpolation is computationally too expensive and other approximate interpolation kernels may need to be used in practice. Moreover, due to finite size of the postage stamp, the signal is never perfectly band-limited and hence is not ideal to begin with. Table~\ref{tab:runtime} shows a significant increase (about 100 fold) in run times when \texttt{sinc} interpolation is used instead of \texttt{Lanczos} kernels.

\begin{figure}
    \centering
    \includegraphics[width=\columnwidth]{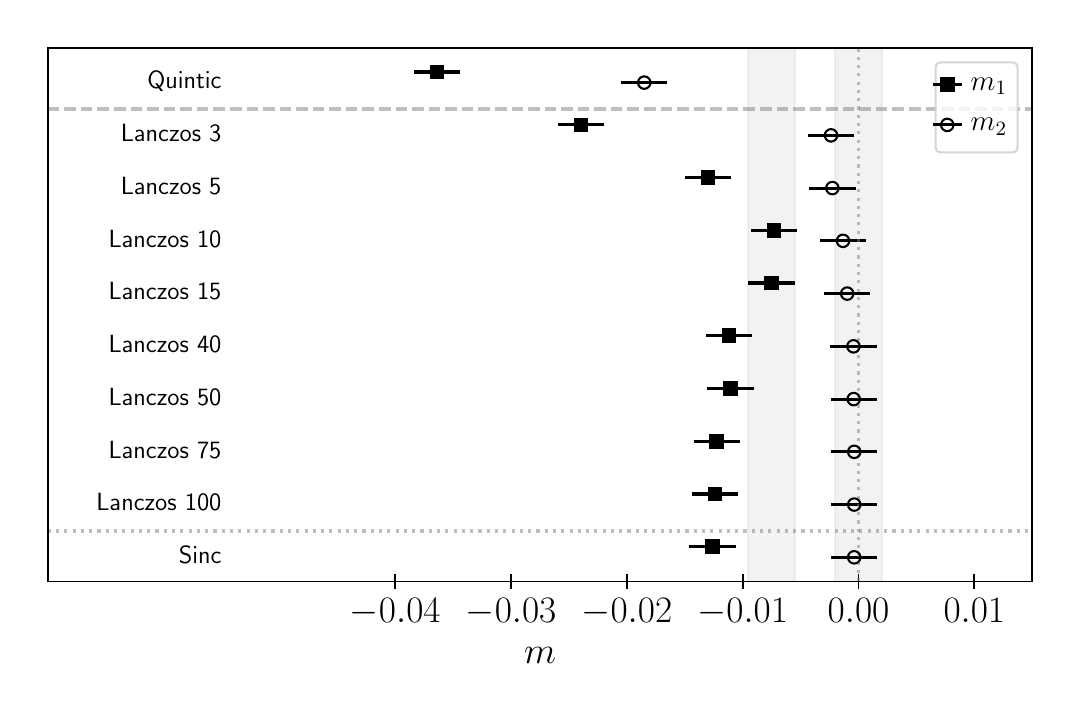}
    \caption{The multiplicative bias $m$ introduced in the shear measured from undersampled images with \metacal\ for \texttt{Quintic}, \texttt{Lanczos}, and \texttt{sinc} interpolation kernels.}
    \label{fig:xinterp}
\end{figure}

For a representative subsample of 500 noiseless galaxy pairs, we show the effect the choice of interpolation kernel has on the measured shear in Fig.~\ref{fig:xinterp}. The \texttt{Quintic} interpolation scheme, which is the default option in \galsim, could potentially introduce biases up to a few percent in both components. However, with the \texttt{sinc} kernel, we find that $m_2$ is fully consistent with zero but $m_1$ shows biases about $-0.01$. We obtain very similar values with \texttt{Lanczos} kernels for very high orders (greater than $75$). For lower order \texttt{Lanczos} kernels, the value of $m_1$ becomes somewhat smaller but still outside the requirements. The decrease in the bias may be attributed to the lower order kernels having a support that is much smaller than typical postage stamp sizes. However, for a {\sc metadetection}-like set-up as in~\cite{Hoekstra2021}, we expect the higher order kernels do perform better. For this reason, we chose to use the $50^\text{th}$ order \texttt{Lanczos} kernel, which is closer to \texttt{sinc} despite it being suboptimal.
\begin{table}
    \centering
    \caption{Time taken for \metacal\ steps per postage stamp.}
    \label{tab:runtime}
    \begin{tabular}{cc}
    \hline
    Interpolation & Average time in seconds \\ 
           kernel & per galaxy per exposure \\
    \hline
\texttt{quintic} & $6.8 \pm 2.0$ \\

\texttt{lanczos3} & $7.0 \pm 1.8$ \\
\texttt{lanczos5}  & $7.3 \pm 1.8$ \\
\texttt{lanczos10} & $6.8 \pm 1.9$ \\
\texttt{lanczos15} & $7.3 \pm 2.5$ \\
\texttt{lanczos50} & $6.4 \pm 1.5$ \\
\texttt{lanczos75} & $9.3 \pm 2.6$ \\
\texttt{lanczos100} & $7.3 \pm 1.9$ \\

\texttt{sinc} & $808.7 \pm 165.4$ \\
\hline
    \end{tabular}
\end{table}

\section{Quadrupole moments in the Fourier domain}
\label{app:quad_moments}
The quadrupole moments defined in equation~(\ref{eq:quad_moments}) may be expressed as
\begin{equation}
    Q_{ij} = -\frac{\partial}{\partial k_i}\frac{\partial}{\partial k_j} \int\rmd^2 \mathbfit{x}\, W(\mathbfit{x})I(\mathbfit{x}) {\rm e}^{-{\rm i}\mathbfit{k}\cdot\mathbfit{x}}\,\biggr\rvert_{\mathbfit{k}=0}.
\end{equation}
The integral is the Fourier transform of $W(\mathbfit{x})I(\mathbfit{x})$ which is given by $(\tilde{W}\otimes\tilde{I})(\mathbfit{k}) := \int\rmd^2\,\mathbfit{k}' \tilde{W}(\mathbfit{k}-\mathbfit{k}')\tilde{I}(\mathbfit{k}')$, where $\tilde{W}$ is the Fourier transform of the weight function and $\tilde{I}$ is the Fourier transform of the image. In terms of the Fourier transforms of the weight functions and images, we can express the quadrupole moments as
\begin{equation}
\begin{split}
       Q_{ij} &= -\frac{\partial}{\partial k_i}\frac{\partial}{\partial k_j} \int\rmd^2\,\mathbfit{k}' \tilde{W}(\mathbfit{k}-\mathbfit{k}')\tilde{I}(\mathbfit{k}')\,\biggr\rvert_{\mathbfit{k}=0} \\
       & = -\int\rmd^2\mathbfit{k}'\, \tilde{W}_{,ij}(-\mathbfit{k}')I(\mathbfit{k}'),
\end{split}
\end{equation}
where $\tilde{W}_{,ij}(\mathbfit{k}) := \frac{\partial}{\partial k_i}\frac{\partial}{\partial k_j}\tilde{W}(\mathbfit{k})$. We deliberately make $\tilde{I}$ a function of only the variable of integration so that the derivatives are taken not on the image $\tilde{I}$ but on the weight function which has an analytic form typically. Using the fact that the weight function is real-valued, we can express $\tilde{W}(-\mathbfit{k}') \equiv \tilde{W}^*(\mathbfit{k}')$, where $*$ denotes complex conjugation. Furthermore, if the weight function has even parity (in real space), $\tilde{W}^*(\mathbfit{k}') \equiv \tilde{W}(\mathbfit{k}')$.
This completes our recasting of the quadrupole moments as integrals in Fourier space.

For Gaussian weight functions, the quadrupole moments may also be expressed as an integral over Fourier modes as
\begin{equation}
       Q_{ij} = -\iint\limits_{|\mathbfit{k}|<k_{\max}} \rmd^2\mathbfit{k} \, q_{ij}(\mathbfit{k}) \tilde{W}(\mathbfit{k})\tilde{I}(\mathbfit{k}),
       \label{eq:quad_moments_F}
\end{equation}
where $q_{ij}(\mathbfit{k})$ is an appropriate quadratic expression, whose details we need not concern ourselves with. The integral is restricted to the disc of radius $k_{\max}$ as $\tilde{I}(\mathbfit{k}) \equiv 0$ outside the disc by definition. We find that the form of the quadrupole moments expressed in equation~(\ref{eq:quad_moments_F}) to be more convenient for this work. Note that this is different from Fourier moments proposed by~\cite{Zhang2008}.

The weighted moments measured post-\metacal\ are
\begin{align}
       Q^\text{interp}_{ij} &= -\iint\limits_{|\mathbfit{k}|<k_{\max}}\rmd^2\mathbfit{k}\, q_{ij}(\mathbfit{k}) \tilde{W}(\mathbfit{k})\tilde{I}^\text{interp}(\mathbfit{k}) \\
                              &= -\sum_{\mathbfit{n}}
                              \iint\limits_{|\mathbfit{k}|<k_{\max}}\rmd^2\mathbfit{k}\, q_{ij}(\mathbfit{k}) \tilde{W}(\mathbfit{k})\tilde{I}(\mathbfit{k}+\mathbfit{n}\Delta_{\mathbfit{k}}),
\end{align}
where $\tilde{I}^\text{interp}({\mathbfit{k}})$ is given in equation~(\ref{eq:I_interp}). When the PSF is undersampled, i.e. $\Delta_{\mathbfit{k}} < 2k_{\max}$, there is an undesirable contribution to $Q^\text{interp}_{ij}$ from some of the $\mathbfit{n} \ne 0$ terms. In particular, for small galaxies, the contribution to $Q^\text{interp}_{ij}$ from $\mathbfit{n}\ne 0$ terms is significant not only due to large amplitude of $\tilde{I}(\mathbfit{k})$ but also due to that of the matched weight function $\tilde{W}(\mathbfit{k})$ for $|\mathbfit{k}|>k_{\max}$. The amplitude of these contributions can be reduced by choosing a weight function that is narrow in Fourier space (equivalently, wide in real space), thereby reducing the amplitude of $q_{ij}(\mathbfit{k})\tilde{W}(\mathbfit{k})$. Thus, the deviation of $Q^\text{interp}_{ij}$ from its true value is small when the weight functions are wide.

\section{Validation on galaxy images with realistic morphologies}
\label{app:morphology}
\begin{figure*}
    \centering
    \includegraphics[width=\textwidth]{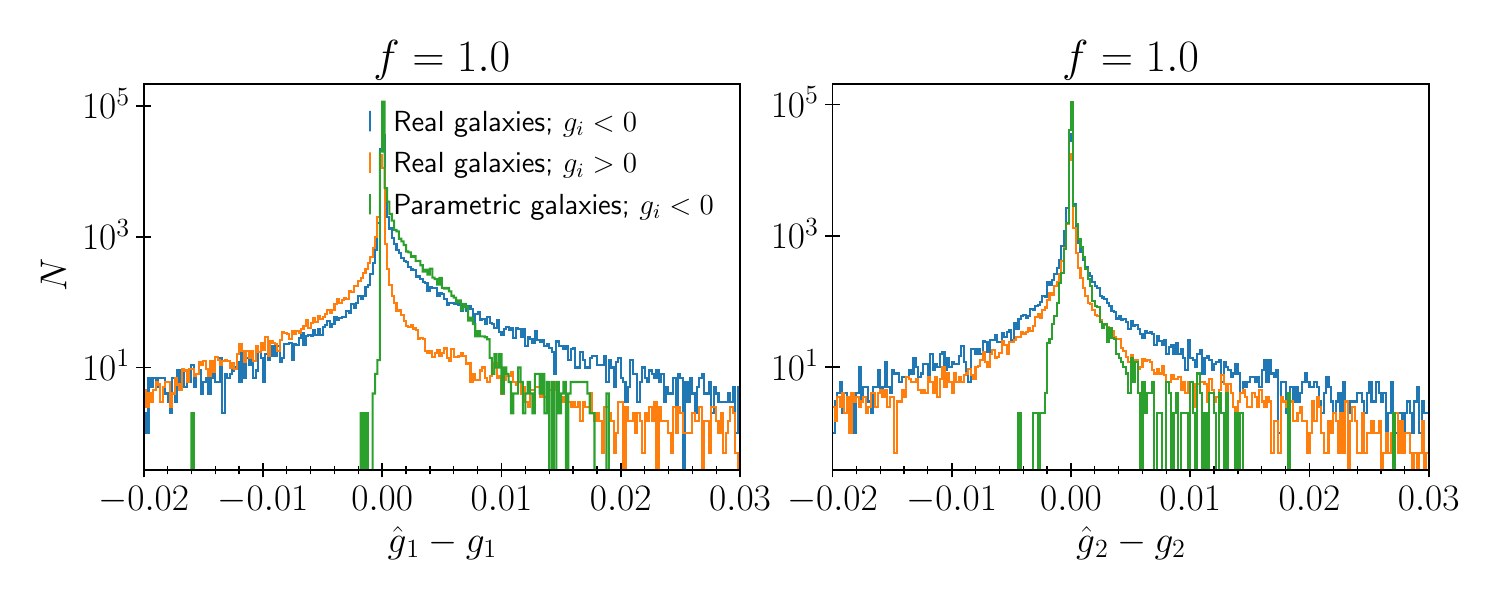}
    \includegraphics[width=\textwidth]{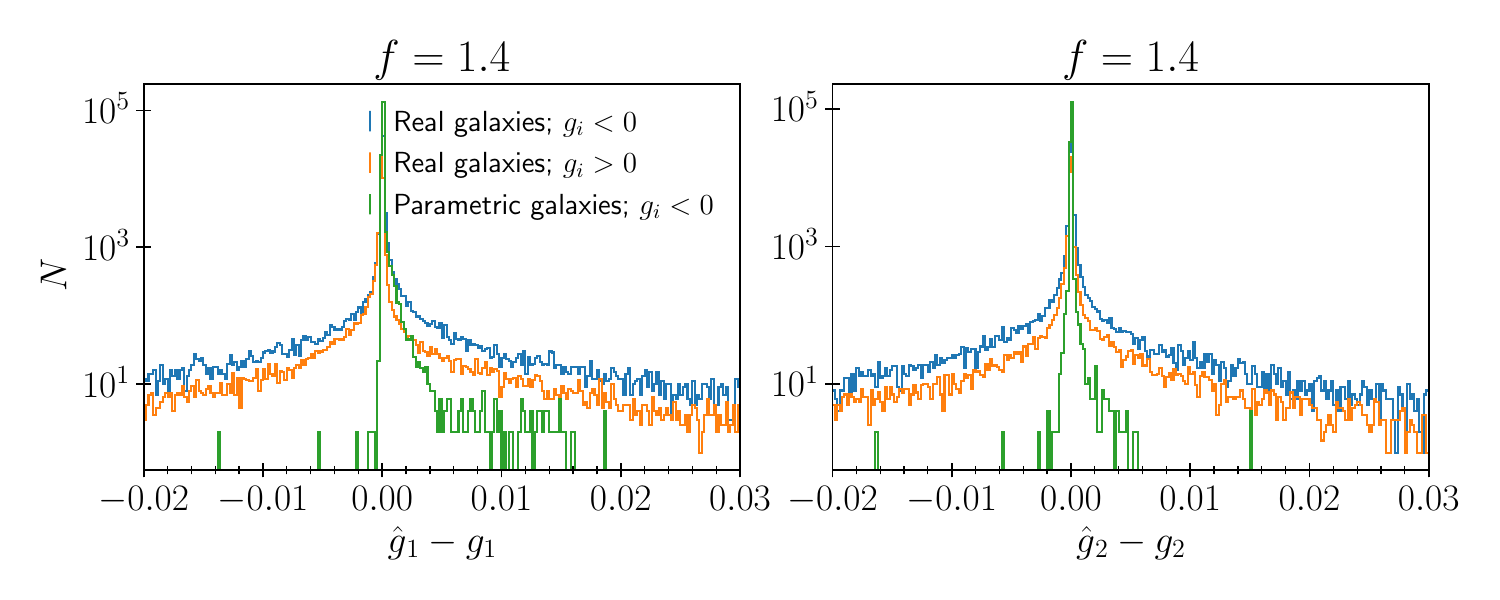}
    \caption{Histogram of shear estimated from pairs of galaxies using the constant scaling method described in Section~\ref{sec:constant_scaling}. The distribution of shear values after subtracting the true shear are shown in blue for real galaxies and in green for parametric galaxies. The skewness towards positive values is due to aliasing bias. To show that the skewness is multiplicative, we include shear estimates for real galaxies with the sign of the true shear values reversed (in orange) and find that the skewness is reversed as well. The lack of parity relation between the blue and orange histograms also indicates the presence of additive bias introduced due to aliasing.  We see in the lower panels that use of a wider weight function mitigates this skewness.}
    \label{fig:rgc}
\end{figure*}

Here, we show that the mitigating strategies discussed in Section~\ref{sec:mitigation} are effective even for galaxies with complex morphologies. Instead of using the best-fitting parameters to the HST COSMOS galaxies, we use the PSF-deconvolved real galaxy images by setting \texttt{real=True} for \texttt{COSMOSCatalog} in \galsim. We refer the reader to~\cite{Mandelbaum2012} for details on how the images were created. We continue to use an Airy PSF with $\lambda_{\text{eff}}=800$\,nm. The remainder of our set-up is the same as described in Section~\ref{sec:simulations}.

We take a different approach in demonstrating that our mitigating strategies work even when the galaxies have realistic morphologies. The measured shears have a much larger fraction of outliers, which prevents us from robustly estimating the values of the shear calibration parameters. We therefore simply show the existence of aliasing bias from the distribution of estimated shear values rather than tabulating the values of the multiplicative bias as before.

We average the shear estimated from each galaxy with its rotated pair to cancel out the intrinsic shape noise and plot the distribution of shear per galaxy pair in Figure~\ref{fig:rgc}. From the upper panels in Figure~\ref{fig:rgc}, we see that for a large majority \sersic\ galaxies, the shear is measured accurately. The one-sided tail, occupied by galaxies with half-light radii smaller than 0\farcs{1}, had introduced the bias in the ensemble shear estimates. In contrast, for realistic galaxies, while the peak is sharply centred at true value, the distribution is two-sided. However, the tail is fatter towards positive values (blue distribution in Figure~\ref{fig:rgc}), thereby biasing the ensemble shear low in amplitude.

To eliminate the possible contribution due to additive biases, we repeated the simulations with the sign of the input shear flipped (orange distribution in Figure~\ref{fig:rgc}). We now find that the tail is fatter towards negative values. This confirms that the ensemble shear is indeed smaller in amplitude relative to the true shear, signalling a negative multiplicative bias as before. The asymmetry between the distributions for positive and negative input shear values strongly suggests the presence of additive bias, which was absent for \sersic\ galaxies. Thus, our single-shear estimator given in equation~\ref{eq:one_point_estimator} cannot be used to evaluate the multiplicative bias.

To demonstrate the efficacy of our mitigation strategy, we repeat the above exercise after scaling the adaptive moments of all galaxies by 1.4. From the lower panels of Figure~\ref{fig:rgc}, it is evident that when we use a larger weight function to estimate the shear,the distribution is much more symmetrical about the true shear value. This is true for both positive and negative values of shear components. Thus, despite the presence of a few outlier measurements, we conclude that the multiplicative bias due to aliasing can be mitigated by adjusting the width of the weight function.


\bsp	
\label{lastpage}
\end{document}